\documentclass[aps, prd, superscriptaddress, nofootinbib, preprintnumbers]{revtex4}
\usepackage[dvipdfmx]{graphicx}
\usepackage{graphicx,bm,color}
\usepackage{subfigure}
\usepackage{amsmath}
\usepackage{amssymb}
\usepackage{amsfonts}
\usepackage{cases}
\usepackage{cancel}
\usepackage[hidelinks]{hyperref}
\usepackage{ulem}
\usepackage{physics}
\usepackage{svg}
\usepackage{natbib}
\usepackage{multirow}

\usepackage{tabularx}

\usepackage{here}
\usepackage{tikz}
\usetikzlibrary{matrix}

\newcommand{\pd}{\partial}
\DeclareMathOperator{\sign}{sign}
\DeclareMathOperator{\Li}{Li}

\begin{document}
	
\title{Bubble wall velocity and gravitational wave in the minimal left-right symmetric model}
	
\author{Dian-Wei Wang}
\email[]{wangdianwei19@mails.ucas.ac.cn}
\affiliation{School of Nuclear Science and Technology, University of Chinese Academy of Sciences, Beijing, P.R.China 100049}
\author{Qi-Shu Yan}
\email[]{yanqishu@ucas.ac.cn}
\affiliation{School of Physics Sciences, University of Chinese Academy of Sciences, Beijing, P.R.China  100049}
\author{Mei Huang}
\email[]{huangmei@ucas.ac.cn}
\affiliation{School of Nuclear Science and Technology, University of Chinese Academy of Sciences, Beijing, P.R.China  100049}

\begin{abstract}
The bubble wall velocity in the first order phase transition plays an important role in determining both the amplitude and the pivot frequency of stochastic gravitational wave background. In the framework of the minimal left-right symmetric model, we study the wall velocity when the first order phase transition can occur. The wall velocity can be determined by matching the distribution functions in the free particle approximation and the local thermal equilibrium approximation. It is found that the wall velocity can be determined in the range $  0.2 <  v_w <  0.5 $ for the parameter space with the first order phase transition. It is also found that for the case when the wall velocity is close to the speed of sound, the peak amplitude of gravitational wave spectrum can be larger than that in the runaway case. Moreover, It is also found that there exists an approximate power law between the wall velocity and pressure difference between broken and symmetry phases, and the power index is equal to 0.41 or so.
\end{abstract}
	
\maketitle
	
\section{Introduction}
\label{sec:intro}
Cosmological  first order phase transition (FOPT) could play some important roles in the baryogenesis, the generation of primordial magnetic field, the formation of primordial black holes, and the production of stochastic gravitational wave backgrounds (SGWB), etc. The production of SGWB during a FOPT could occur via three major processes: bubble collisions \cite{cite:bubblecol1, cite:bubblecol2, cite:bubblecol3, cite:bubblecol4, cite:bubblecol5, cite:bubblecol6}, sound wave \cite{cite:SW1, cite:SW2, cite:SW3, cite:Hindmarsh15} and megnatohydrodynamic (MHD) turbulence\cite{cite:MHD1, cite:MHD2, cite:MHD3, cite:MHD4, cite:MHD5}. The GWs produced by a FOPT could be detected by the proposed GW experiments, including TianQin \cite{cite:TianQin}, Taiji \cite{cite:Taiji}, LISA \cite{cite:LISA1, cite:LISA2}, ALIA \cite{cite:ALIA}, MAGIS \cite{cite:MAGIS}, DECIGO \cite{cite:DECIGO}, BBO \cite{cite:BBO}, Cosmic Explore (CE) \cite{cite:CE}, Einstein Telescope (ET) \cite{cite:ET}, and aLIGO \cite{cite:aLIGO}. A pedagogical review from the finding FOPT in a particle physics model to predicting GWs can be found in \cite{cite:Athron23}. 
	
Typically, a cosmological FOPT can be initiated by the stochastic nucleation of vacuum bubble. A scalar field, as an order parameter, can develop a non-vanishing VEV inside the bubble, while its value in the false vacuum vanishes. The effective potential energy inside the bubble is lower than that of outside. While the pressure inside the bubble is larger than that of outside. The direction of the net pressure is pointed outside, which acts as a driving force on the bubble wall and pushes the bubble to expand. When the bubble wall starts to move, plasma in the vicinity of the wall exert a backreaction (sometime is also called as friction force) on the wall. If such a friction force can balance the net pressure, the bubble wall can reach to its terminal velocity $ v_w $. 
	
On the first hand, in the process that the wall sweeps through the plasma, the deviations from the equilibrium in vicinity of the wall can be produced due to the varying of scalar field. The deviation from equilibrium state of the plasma consequently affect the determination of $ v_w $ through its modification in friction force on the bubble wall. On the other hand, at a length scale much larger than the width of wall and a time scale larger than the relaxation time, the deviations have dissipated and the plasma behind the wall can reach to its new thermal equilibrium. Thus hydrodynamics can be used to describe the bulk motion of the plasma. In the radiation dominance epoch, the bulk kinetic energy of the plasma dominates the sources \cite{cite:spectrum} of the gravitational wave produced in FOPT. 
	
In the study of bubble wall dynamics, determining $ v_w $, or equivalently the relation between friction force versus $ v_w $, had been a challenge. To consider the interaction between scalar field and the plasma, together with collision between particles, one need to set up transport equations of particles, \cite{cite:Linde92EWPT, cite:BHLiu92, cite:MoPr95PRL, cite:MoPr95}. For a bubble wall with $ \gamma_w \gtrsim 1 $ ($\gamma_w$ is the boost factor of bubble wall in the bubble center frame), the moment method and expansion near equilibrium can be used to solve the Boltzmann equation \cite{cite:Ko14, cite:Cline20singlet}. Different parameterization of the deviations from equilibrium has been discussed in recent literature \cite{cite:newform, cite:sonicboom, cite:newnewform}. Determining $ v_w $ with barely hydrodynamic information has also been studied in local thermal equilibrium (LTE) limit \cite{cite:AWYresult, cite:WSJresult, cite:AWYcrit}. In contrast, for very relativistic bubble wall with $ \gamma_w \gg 1 $, it is found that the emission of soft gauge bosons during a particle passing through the wall dominates the friction force \cite{cite:gamma0, cite:gamma1, cite:gamma2, cite:gamma1g2, cite:Azatov20, cite:Azatov23}. A phenomenological friction coefficient $ \eta $ can be added into the scalar field equation, and the relation between $ \eta $ and $ v_w $ has been studied in literature \cite{cite:ErgBdg, cite:Huber11, cite:Megevand}.
	
In order to measure how much the vacuum energy has been converted into the bulk kinetic energy, we need to determine the efficiency factor $ \kappa_v $. The energy budget of an expanding bubble has been studied in Reference \cite{cite:ErgBdg}, where  a bag equation of state (EoS) with $ c_s^2 = 1 / 3 $ is assumed in both symmetric phase and broken phase. For more general cases of EoS with $ c_{s,s}^2 $ and $ c_{s,b}^2 \neq 1 / 3 $ but constant have been studied in \cite{cite:ErgBdgConstcs, cite:HFPeng21beyondbag, cite:Ko20, cite:Ko21}. By expanding the effective potential with respect to the temperature, the authors of the reference \cite{cite:beyondbag} considered a model independent method by including the temperature dependency of $ c_s^2 $. 
	
In this work, to determine the bubble wall velocity, we used two approaches that works in two different length scales. At a length scale which is much less than the mean free path, we use the free particle approximation to describe the momentum transmission between particles and the bubble wall. At the length scale which is much greater than the mean free path, we adopt the local thermal equilibrium to describe the fluid bulk motion. In order to determine the bubble wall velocity, we demand these two distribution functions can have the same momentum flux. 

It is well-known that the electroweak symmetry breaking of the standard model (SM) is a crossover and there is no FOPT. In order to study the bubble wall dynamics concretely, we choose the minimal left-right symmetric model (MLRSM) \cite{cite:LRSM1, cite:LRSM2, cite:LRSM3} as the underlying particle physics model, where the first order phase transitions can occur when the neutral component of the right-hand triplet is light enough (saying a few hundred GeV) as found in the Reference \cite{cite:mingqiu}. The MLRSM enlarges the EW sector of SM $ SU(2)_L \times U(1)_Y $ to $ SU(2)_R \times SU(2)_L \times U(1)_{B - L} $, and brings new insight to important questions of the SM, like spontaneous parity violation/restoration, CP violation, and neutrino masses. 

In the Reference \cite{cite:mingqiu}, the bubble wall velocity is taken as $v_w=1$. To our best knowledge, in the context of the MLRSM, there is no work yet on how to determine the bubble wall velocity dynamically. This work is supposed to fill this gap and we will study how the bubble wall velocity in the FOPT of the MLRSM can be determined by the phase transition dynamics and how the gravitational waves spectra can be modified when compared with the case $v_w=1$.

The paper is organized as follows. In section \ref{sec:lrsm and pt}, we briefly review the FOPT in the MLRSM. In section \ref{sec:bubble dynamics}, we derive the formula that describe the dynamics in field-plasma system, and compute the bubble wall velocity. In section \ref{sec:GW}, we examine the effects of bubble wall velocity to the gravitational wave produced from the FOPT. In section \ref{sec:dnc}, we end this work with some discussions and conclusions.
	
\section{The First order Phase transition in the MLRSM}
\label{sec:lrsm and pt}
\subsection{A brief review on the Higgs potential of the MLRSM}
The MLRSM enlarges the electroweak sector $ SU(2)_L \times U(1)_Y $ in SM gauge group to $ SU(2)_R \times SU(2)_L \times U(1)_{B-L} $, and the fermions are arranged in an apparent left-right symmetric fashion. In the MLRSM, the Higgs sector include a Higgs bi-doublet $ \Phi $, a left-handed triplet $ \Delta_L $ and a right-handed triplet $ \Delta_R $
\begin{equation}
\Phi = \mqty(\phi_1^0 & \phi_2^+ \\ \phi_1^- & \phi_2^0),\quad
		\Delta_L = \mqty(\Delta_L^+ / \sqrt2 & \Delta_L^{++} \\ \Delta_L^0 & -\Delta_L^+ / \sqrt2),\quad
		\Delta_R = \mqty(\Delta_R^+ / \sqrt2 & \Delta_R^{++} \\ \Delta_R^0 & -\Delta_R^+ / \sqrt2).
\end{equation}
	
The most general scalar potential in the LRSM can be written as
\begin{eqnarray}
\label{eqn:potential}
	{\cal V} &=& -\mu_{1}^{2} \operatorname{Tr}[\Phi^{\dagger} \Phi]-\mu_{2}^{2}\left(\operatorname{Tr}[\tilde{\Phi}
	\Phi^{\dagger}]+\operatorname{Tr}[\tilde{\Phi}^{\dagger} \Phi]\right)-\mu_{3}^{2}\left(\operatorname{Tr}[\Delta_{L}  \Delta_{L}^{\dagger}]+\operatorname{Tr}[\Delta_{R} \Delta_{R}^{\dagger}]\right) \nonumber \\
	  && +\rho_{1}\left(\operatorname{Tr}[\Delta_{L} \Delta_{L}^{\dagger}]^{2}+\operatorname{Tr}[\Delta_{R} \Delta_{R}^{\dagger}]^{2}\right)+\rho_{2}\left(\operatorname{Tr}[\Delta_{L} \Delta_{L}] \operatorname{Tr}[\Delta_{L}^{\dagger} \Delta_{L}^{\dagger}]+\operatorname{Tr}[\Delta_{R} \Delta_{R}] \operatorname{Tr}[\Delta_{R}^{\dagger} \Delta_{R}^{\dagger}]\right) \nonumber \\
	  && +\rho_{3} \operatorname{Tr}[\Delta_{L} \Delta_{L}^{\dagger}] \operatorname{Tr}[\Delta_{R} \Delta_{R}^{\dagger}]+\rho_{4}\left(\operatorname{Tr}[\Delta_{L} \Delta_{L}] \operatorname{Tr}[\Delta_{R}^{\dagger} \Delta_{R}^{\dagger}]+\operatorname{Tr}[\Delta_{L}^{\dagger} \Delta_{L}^{\dagger}] \operatorname{Tr}[\Delta_{R} \Delta_{R}]\right) \nonumber \\
	  && +\lambda_{1} \operatorname{Tr}[\Phi^{\dagger} \Phi]^{2}+\lambda_{2}\left(\operatorname{Tr}[\tilde{\Phi} \Phi^{\dagger}]^{2}+\operatorname{Tr}[\tilde{\Phi}^{\dagger} \Phi]^{2}\right) \nonumber \\
	  && +\lambda_{3} \operatorname{Tr}[\tilde{\Phi} \Phi^{\dagger}] \operatorname{Tr}[\tilde{\Phi}^{\dagger} \Phi]+\lambda_{4} \operatorname{Tr}[\Phi^{\dagger} \Phi]\left(\operatorname{Tr}[\tilde{\Phi} \Phi^{\dagger}]+\operatorname{Tr}[\tilde{\Phi}^{\dagger} \Phi]\right)\nonumber \\
	  && +\alpha_{1} \operatorname{Tr}[\Phi^{\dagger} \Phi]\left(\operatorname{Tr}[\Delta_{L} \Delta_{L}^{\dagger}]+\operatorname{Tr}[\Delta_{R} \Delta_{R}^{\dagger}]\right)+\alpha_{3}\left(\operatorname{Tr}[\Phi \Phi^{\dagger} \Delta_{L} \Delta_{L}^{\dagger}]+\operatorname{Tr}[\Phi^{\dagger} \Phi \Delta_{R} \Delta_{R}^{\dagger}]\right)\nonumber \\
	  && +\left[ \alpha_{2} e^{i \delta} \left(\operatorname{Tr}[\Delta_{L} \Delta_{L}^{\dagger}] \operatorname{Tr}[\tilde{\Phi} \Phi^{\dagger}]+\operatorname{Tr}[\Delta_{R} \Delta_{R}^{\dagger}] \operatorname{Tr}[\tilde{\Phi}^{\dagger} \Phi] \right )+\mathrm{H.c.} \right] \nonumber \\
	  && +\beta_{1}\left(\operatorname{Tr}[\Phi \Delta_{R} \Phi^{\dagger} \Delta_{L}^{\dagger}]+\operatorname{Tr}[\Phi^{\dagger} \Delta_{L} \Phi \Delta_{R}^{\dagger}]\right)+\beta_{2}\left(\operatorname{Tr}[\tilde{\Phi} \Delta_{R} \Phi^{\dagger} \Delta_{L}^{\dagger}]+\operatorname{Tr}[\tilde{\Phi}^{\dagger} \Delta_{L} \Phi \Delta_{R}^{\dagger}]\right) \nonumber \\
	  && +\beta_{3}\left(\operatorname{Tr}[\Phi \Delta_{R} \tilde{\Phi}^{\dagger} \Delta_{L}^{\dagger}]+\operatorname{Tr}[\Phi^{\dagger} \Delta_{L} \tilde{\Phi} \Delta_{R}^{\dagger}]\right), \label{pot}
\end{eqnarray}
where $\tilde{\Phi} = \sigma_2 \Phi^\ast \sigma_2$ (with $\sigma_2$ the second Pauli matrix). As being required by left-right symmetry, all the quartic couplings in the potential above are real parameters. The CP violating phase $\delta$ associated with $\alpha_2$ is shown explicitly.

In this work, to simplify our discussion on the velocity of bubble wall, we confine to consider a specific scenario where there exists FOPT which is mainly driven by the neutral component of $\Delta_R$ and the phase transition occurs to break the symmetry $SU_R(2) \times U_{B-L}(1) \to U_Y(1)$, as pointed out in \cite{cite:mingqiu}. After taking into account current experimental bounds and theoretical constraints, the scenario can set parameters as 
$\rho_4 = \lambda_2  = \lambda_3 =\lambda_4 = \alpha_1 =\alpha_2 = \beta_1 = \beta_2 = \beta_3 = 0$.
	
	At zero temperature, the neutral components of scalar fields can develop non-zero vaccuum expectation values, which can be parameterized as follows
	\begin{equation}
		\expval{\Phi} = \frac{1}{\sqrt2} \mqty(\kappa_1 & 0 \\ 0 & \kappa_2 e^{i \theta_\kappa}),\quad
		\expval{\Delta_L} = \frac{1}{\sqrt2} \mqty(0 & 0 \\ v_L e^{i \theta_L} & 0),\quad
		\expval{\Delta_R} = \frac{1}{\sqrt2} \mqty(0 & 0 \\ v_R & 0),\qq{with}
		T = 0,
	\end{equation}
where, $ \theta_{L, \kappa} $ are CP violating phases and set to be zero in this work. While the bi-doublet VEV $ \kappa_{1, 2} $ are related to EW VEV through the relation $ v_\mathrm{EW} = \sqrt{\kappa_1^2 + \kappa_2^2} = 246 \, \mathrm{GeV} $. And the ratio between $ \kappa_1 $ and $ \kappa_2 $ is set to be $ \xi = \kappa_2 / \kappa_1 = 10^{-3} $, as required by experimental constraints. 
	
Due to the tiny neutrino mass, it is a good approximation to take $ v_L = 0 $. Thus, In this simplified scenario, the theoretical free parameters of the potential can be traded off by physical particle masses as given follows
	\begin{equation}
		\begin{split}
			\rho_1 &= \frac{m_{H_3^0}^2}{2 v_R^2},\quad
			\rho_2 = \frac{m_{H_3^{++}}^2}{2 v_R^2},\quad
			\rho_3 = \frac{2 m_{H_2^0}^2 - m_{H_3^0}^2}{v_R^2},\quad
			\alpha_3 = \frac{2 m_{H_3^0}^2}{v_R^2}\\
			\frac{\mu_1^2}{v_R^2} &= - \frac{\alpha_3}{2} \frac{\xi^2}{1 - \xi^2} + \lambda_1 \frac{v_\mathrm{EW}^2}{v_R^2},\quad
			\frac{\mu_2^2}{v_R^2} = \frac{\alpha_3}{4} \frac{\xi^2}{1 - \xi^2},\quad
			\frac{\mu_3^2}{v_R^2} = \rho_1  + \frac{\alpha_3}{2} \frac{v_\mathrm{EW}^2}{v_R^2}. 
		\end{split} \label{msp}
	\end{equation}
Due to the left-right symmetry, both gauge coupling constants of $ SU(2)_{L, R} $ are fixed to be $ g_L = g_R = 0.64 $, and the gauge coupling constant of $U_{B-L}(1)$ denoted as $ g_{B - L} $ is determined by the following relation 
	\begin{equation}
		\frac{1}{e^2} = \frac{1}{g_L^2} + \frac{1}{g_Y^2} = \frac{1}{g_L^2} + \frac{1}{g_R^2} + \frac{1}{g_{B - L}^2}.
	\end{equation}
To simplify our study, we assume that the symmetry breaking is $ SU(2)_R \times U_{B-L}(1) \to U_Y(1)$, of which the breaking scale is determined by  $ v_R $ (which is typically 10 TeV or above).  Thus, in this approximation, the VEV of the triplet scalar $ \Delta_R $ becomes non-vanishing, while the VEV of other scalar fields are set to be vanishing
	\begin{equation}
		\expval{\Phi} = \expval{\Delta_L} = 0,\quad
		\expval{\Delta_R} = \frac{1}{\sqrt2} \mqty(0 & 0 \\ \phi & 0),\qq{with}
		T \sim v_R.
	\end{equation}
where $\phi$ is the neutral component of 	$\Delta_R$, which plays the key role to determine the phase transition.
	
\subsection{The first order Phase transition}
We start from the effective potential, which is a function of $\phi$ and can be put as given below
	\begin{equation}
	\label{eq:Veff}
		\begin{split}
			V_\mathrm{eff}
			&= V_0(\phi) + V_1^{T = 0}(\phi) + V_1^{T \neq 0}(T, \phi)\\
			&= \qty(- \frac{\mu_3^2}{2} \phi^2 + \frac{\rho_1}{4} \phi^4) + \frac{1}{64 \pi^2} \sum_i g_i m_i^4 \qty(\log \frac{m_i^2}{\mu^2} - C_i) + \frac{T^4}{2 \pi^2} \sum_i g_i J_\pm \qty(\frac{m_i^2}{T^2}),
		\end{split}
	\end{equation}
where $V_0$ is the tree-level potential, $ V_1^{T = 0} $ is the Coleman-Weinberg one-loop effective potential, and $ V_1^{T \neq 0} $ is thermal contribution at finite temperature. 
	
The mass parameters $m_i$ of different particles are dependent on the value of $\phi$, as given in Eq. (\ref{msp}) . The renormalization scale $ \mu $ is tuned to make the location of minimum of $ V_\mathrm{eff} $ remains $ \phi = v_R $. The degree of freedom $ g_i $ and constant $ C_i $ for $ i $-th particle are given by
	\begin{equation}
		(g_i, C_i) = 
		\begin{cases}
			(1, 3 / 2), & \text{for scalar,}\\
			(-2 \lambda, 3 / 2), & \text{for fermion,}\\
			(3, 5 / 6), & \text{for gauge boson,}
		\end{cases}
	\end{equation}
where $ \lambda = 1 \ (2)$ for Weyl (Dirac) fermion. In the bosonic sector, there are Higgs scalars and vector gauge bosons which can contribute to this Higgs potential, and their contributions are crucial for the FOPT. In this study, the Goldstone bosons are treated as the longitudinal components of vector gauge bosons. In the fermionic sector, there are right-handed neutrino which can contribute significantly to this Higgs potential.

It is noteworthy that in order to get rid of imaginary part in $ V_\mathrm{eff} $ that arises from $ - \mu_{1,2,3}^2 $ in scalar scalar masses, we use thermal mass $ m_i^2 = m_i^2(T, \phi) $ instead of $ m_i^2(0, \phi) $ for every scalars, this also recovers the correction $ V_D $ from daisy diagram to the effective potential $ V_\mathrm{eff} $ in previous study\cite{cite:mingqiu}, the actual formulae for particle masses are listed in Appendix \ref{sec:mass}. For particles in thermal equilibrium, function $ J_- \ (J_+) $ for bosons (fermions) are defined as
	\begin{equation}
		J_\pm(m^2 / T^2) = T^{-3} \int_0^\infty \dd{p} p^2 \log(1 \pm e^{-\sqrt{p^2 + m^2} / T}).
	\end{equation}
	
When the FOPT happens, a critical bubble can form through thermally assisted quantum tunneling process. The nucleation rate of bubbles is given by \cite{cite:S3bounce}
	\begin{equation}
		\Gamma = \Gamma_0 \exp(-S_3 / T), \quad
		\Gamma_0 \sim T^4 (S_3 / 2 \pi T)^{3 / 2},
	\end{equation}
	where $ S_3 $ is 3D tunneling action, which is given as
	\begin{equation}
		S_3 = \int \dd[3]{x} \qty[\frac{1}{2} (\grad \phi_\mathrm{b})^2 + \Delta V_\mathrm{eff}(T, \phi_\mathrm{b})],
	\end{equation}
	where $ \Delta V_\mathrm{eff}(T, \phi) $ is defined as $ \Delta V_\mathrm{eff}(T, \phi) = V_\mathrm{eff}(T, \phi) - V_\mathrm{eff}(T, 0) $. And the field configure $ \phi_\mathrm{b} $ is a saddle point of $ S_3 $ known as bounce solutions which satisfy the following equation of motion and boundary conditions
	\begin{equation}
		\pdv[2]{\phi}{r} + \frac{2}{r} \pdv{\phi}{r} = \pdv{v} V_\mathrm{eff}(T, \phi), \quad
		\phi'(r = 0) = 0, \quad
		\phi(r \to \infty) = 0.
	\end{equation}
	In this study, we use the code  \texttt{CosmoTransitions} \cite{cite:CosmoTransitions} to find the bounce solution and to evaluate $ S_3 $ when $\phi_b$ is found.  The nucleation temperature $ T_N $ is defined when $ \Gamma / H \sim \order{1} $, where the Hubble's parameter is defined as 
	\begin{equation}
		H_*(T) = \frac{8 \pi^3 g_* T^2}{90 M_\mathrm{pl}},
	\end{equation}
	with $ M_\mathrm{pl} $ denoting the Planck mass. The total degree of freedoms of the model can be found to be $ g_* = 134 $.

	In order to determine the GW spectrum caused by the FOPT, there are two parameters needed to be obtained. The first one is $\alpha$, which describes the strength of the phase transition
	\begin{equation}
		\alpha = \eval{\frac{\epsilon_*(T)}{a_+(T) T^4}}_{T = T_N},
	\end{equation}
	where
	\begin{equation}
		\epsilon_* = \eval{- \Delta V_\mathrm{eff} + \frac{T}{4} \frac{\pd \Delta V_\mathrm{eff}}{\pd T}}_{T = T_N}, \qq{and}
		a_+(T) T^4 = - \eval{\frac{3 T}{4} \pdv{V_\mathrm{eff}(T, 0)}{T}}_{T = T_N}.
	\end{equation}
	The other one is $\beta$, which describes the inverse duration of the phase transition compared with Hubble's parameter
	\begin{equation}
		\frac{\beta}{H_*} = \eval{T \dv{T} \frac{S_3}{T}}_{T = T_N}.
	\end{equation}
	\begin{figure}[h!]
		\centering
		\includegraphics[width = 0.7\columnwidth]{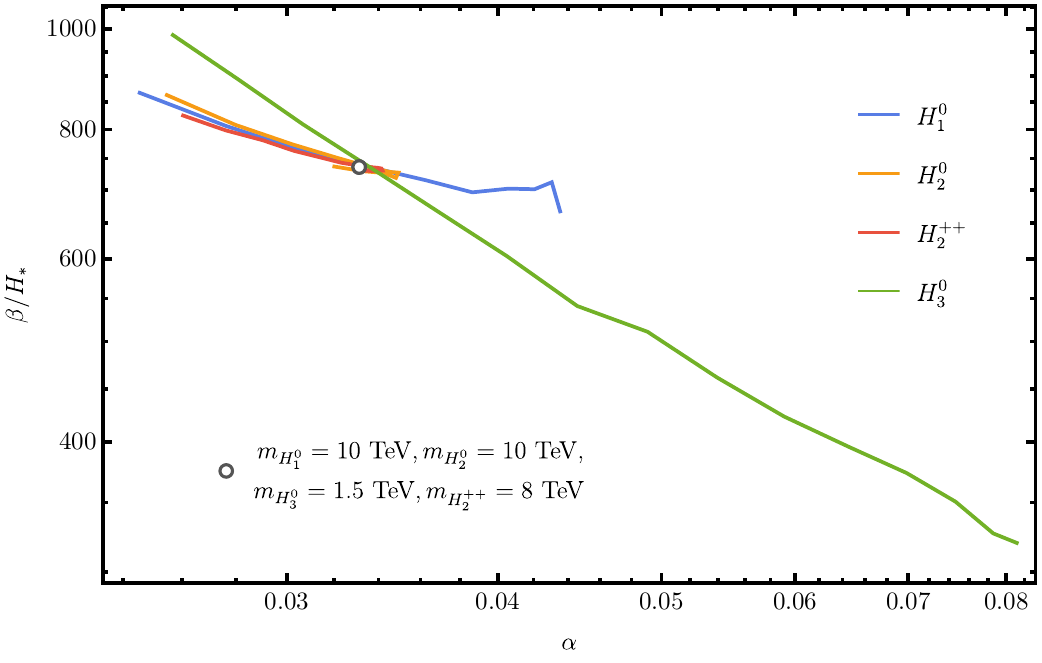}
		\caption{Model parameter dependence of two feature parameters of phase transition, $\alpha $  and $ \beta / H_* $, are shown.}
		\label{fig:alphabeta}
	\end{figure}
	
Below we demonstrate the effects of model parameters of Higgs potential to these two parameters. As mentioned above, the model parameters of Higgs potential can be expressed as the masses of scalars, i.e. $ m_{H_1^0} $, $ m_{H_2^0} $, $ m_{H_2^{++}} $ and $ m_{H_3^0} $ separately. The mass ranges for these scalar particles are tabulated in the last two columns in \tablename{} \ref{tab:sclmass}, which are consistent with both experimental bounds  and theoretical constraints, which had been studied in reference \cite{cite:mingqiu}. 

The degeneracy of each parameter is also tabulated  in \tablename{} \ref{tab:sclmass}, and the detailed mass relations in the thermal media are provided in the Appendix \ref{sec:mass}. For example, there are four scalar particles which have the same mass, i.e. $H_1^0$, $A_1^0$, and $H_1^\pm$, as given in Eq. (\ref{massrel}), and the mass parameter is denoted by $m_{H_1^0}$.
\begin{table}[h!]
\centering
\caption{Default values and range of scalar particle masses are tabulated and the degeneracy of each scalar mass is also provided.}
\label{tab:sclmass}
\begin{tabularx}{0.8\textwidth}
                {>{\centering\arraybackslash}X
				>{\centering\arraybackslash}X
				>{\centering\arraybackslash}X
				>{\centering\arraybackslash}X
				>{\centering\arraybackslash}X}
			\hline\hline
			parameters & degeneracy & default value & lower bound & upper bound \\
			\hline
			$ m_{H_1^0} $ & 4 & 10 TeV & 6 TeV & 12 TeV \\
			\hline
			$ m_{H_2^0} $ & 6 & 10 TeV & 4 TeV & 10.5 TeV \\
			\hline
			$ m_{H_2^{++}} $ & 2 & 8 TeV & 4 TeV & 11.5 TeV \\
			\hline
			$ m_{H_3^0} $ & 1 & 1.5 TeV & 0.125 TeV & 2 TeV \\
			\hline\hline
		\end{tabularx}
	\end{table}
	
Based on the input of the model given in \tablename{} \ref{tab:sclmass}, in \figurename{} \ref{fig:alphabeta}, we show the computed parameters $ \alpha $ and $ \beta / H_* $ with different particle masses.  A set of default values for the masses of scalars are given as an open circle, while the range of mass values of scalars are listed in the second column in \tablename{} \ref{tab:sclmass}. For each curve, we vary the mass of one particle, and fix the masses of other particles to their default values. For example, for the blue line, we vary the mass of $H_1^0$ from $6-12$ TeV, and fix other particles' masses at the default values.
		
From \figurename{} \ref{fig:alphabeta}, it is observed that in general, $ \beta / H_* $ is inversely proportional to $ \alpha$.  The masses of scalars $H_1^0$, $H_2^0 $ and $H_2^{++}$ are not sensitive to the results, by varying their masses, the value of $ \alpha$ is around 0.03 and $ \beta / H_* $ is around 700. However, the mass of scalar $ H_3^0 $  can much more drastically change the values of both $ \alpha $  and $ \beta / H_* $, especially $ \beta / H_* $ can reach 200 at $ \alpha $ around 0.08. This can be attributed to the fact that $ m^2_{H_3^0} = 2 \rho_1 v_R^2 $, where $ \rho_1 $ directly contributes to $ V_\mathrm{eff} $ through the quartic term at the tree-level, while other scalar particles only contribute to the effective potential through one-loop effects and thermal loops. It is worthy of mentioning that, the calculation of the left-right model in \cite{cite:mingqiu} showed that $\beta / H_* $ can reach $10^4$ and $\alpha$ can be as small as 0.01, which are dependent upon the theoretical parameters. 
	
\section{Dynamics for steady bubble wall}
\label{sec:bubble dynamics}
When a bubble forms, the vacuum pressure difference between the inside and the outside pushes the bubble to expand, i.e. $\Delta p = p_i - p_o>0$, where $p_i$ and $p_o$ denotes the pressure inside and outside of the bubble, respectively. The moving bubble wall stirs the ambient plasma, pushing the plasma out of equilibrium, and the plasma exerts a friction force on the bubble wall. The friction force increases with the bubble wall velocity. If the friction force on per unit area on the bubble wall balances the vacuum pressure difference before the bubble wall collision, the wall will reach its terminal velocity $ v_w $, otherwise the wall will run away, i.e. the Lorentz factor $ \gamma_w $ increase permanently. The friction force in high $ \gamma_w $ limit has been studied in \cite{cite:gamma0, cite:gamma1, cite:gamma2, cite:gamma1g2}, where the particle splitting process dominates the friction force. In this work, it is found that the friction is strong enough in $ \gamma_w \gtrsim 1 $ region and the typical value for the bubble velocity $ v_w $ is $ 0.2 - 0.5 $ , thus we can neglect particle splitting  safely.

	\subsection{The field-plasma system}
	\label{subsec:fieldplasma}
	The energy-momentum tensor of the field-plasma system can be split in two parts, i.e. the scalar field part $T^{\mu\nu}_\mathrm{field}$ and the plasma part $T^{\mu\nu}_\mathrm{plasma}$. And the conservation law can be expressed as 
	\begin{equation}
		\label{eq:conserve}
		\pd_\mu T^{\mu\nu} = \pd_\mu T^{\mu\nu}_\mathrm{field} + \pd_\mu T^{\mu\nu}_\mathrm{plasma} = 0.
	\end{equation}
	The energy-momentum tensor of scalar field $T^{\mu\nu}_\mathrm{field}$ is given by
	\begin{equation}
		\label{eq:TmunuField}
		T^{\mu\nu}_\mathrm{field} = \pd^\mu \phi \pd^\nu \phi - g^{\mu\nu} \qty(\frac{1}{2} \pd_\rho \phi \pd^\rho \phi - V_\mathrm{vac}(\phi)).
	\end{equation}
	where $ V_\mathrm{vac}(\phi) $ is the renormalized vacuum effective potential. The energy-momentum tensor of plasma $T^{\mu\nu}_\mathrm{plasma}$ is given by
	\begin{equation}
		\label{eq:TmunuPlasma}
		T^{\mu\nu}_\mathrm{plasma} = \sum_i \int \frac{\dd[3]p}{(2 \pi)^3 E_i} p^\mu p^\nu f_i(\vb p, \vb x).
	\end{equation}
	where $ i $ runs over the particle species. In order to study the dynamics at the microscopic scale, we neglect the curvature of the bubble wall and assume that the wall can be described by two parallel planes being separated by the distance $L_w$ and both moving towards $z $-direction.  For the non-runaway case, the bubble wall velocity $ v_w $ reach its terminal value before collision, we have
	\begin{equation}
		\label{eq:ansatzSteady}
		\phi = \phi\qty( \frac{\bar u_w^\mu x_\mu}{L_w}),
	\end{equation}
	where $ \bar u_w^\mu = \gamma_w (v_w, 0, 0, 1) $ is a space-like unit vector and $ L_w $ is the bubble wall thickness. 
	
	The dynamical transition of particles in the plasma can be described by the Boltzmann equation
	\begin{equation}
		\label{eq:Boltzmann}
		L[f] \equiv \qty(\pdv{t} - \frac{p_z}{E} \pdv{z} + \frac{m}{E} \pdv{m}{z} \pdv{p_z}) f(\vb p, \vb x) = C[f],
	\end{equation}
	where $ L $ is an abbreviation of the differential operator on the lhs, $f(\vb p, \vb x)$ is the distribution functions of a particle in the phase space, and $ C[f] $ is collision integral.
	
In earlier works, to determine the bubble velocity, the Boltzmann equations are solved using the moment method \cite{cite:MoPr95}, or the spectral method \cite{cite:newnewform}. Both methods need to solve the coupled differential equations numerically for both active and passive particles, which becomes an obstacle when the species of particles become larger like in the left-right symmetric model where there are more than 10 types of active particles which should be taken into account. For example, there are 3 family right-handed neutrinos, 3 massive gauge vector particles (each has 3 degree of freedom), and 4 types of scalar fields with 13 degrees of freedom as given in Table \ref{tab:sclmass}. 

In order to circumvent this obstacle, we will adopt two methods to approximate the field-plasma system, i.e. 1) free particle approximation and 2) local thermal equilibrium (LTE) approximation. We match these two approximations to find the bubble velocity, As we know, these two approaches work at different length scale. 

\begin{itemize}
\item At the length scale where the distance to bubble wall is much smaller than the mean free paths $ \lambda_\mathrm{mfp} $, the $ \pd_z m $ term on lhs of Eq. \eqref{eq:Boltzmann} dominates, thus the collision terms $ C[f] $ can be neglected. At this distance, the free particle approximation can be used, and there is an analytical solution to Eq. \eqref{eq:Boltzmann}, which is discussed in detail in \ref{subsec:freeparticle}. We use this approximation to estimate the thermal pressure, which acting as a friction force on the bubble wall. 

\item In another limit where the distance much lager than $ \lambda_\mathrm{mfp} $, LTE becomes a good approximation. We utilize this method to calculate the profile of sound shell, and more importantly, to calculate energy and momentum flux near the bubble wall, which is discussed in \ref{subsec:LTE}. 

\end{itemize}
The key criteria is to compute the mean free paths $ \lambda_\mathrm{mfp} $. Considering a typical scattering process $ 12 \to 34 $ which occurs near thermal equilibrium, the mean free paths can be calculated from the collision terms as 
	\begin{equation}
	\begin{split}
	    \lambda_\mathrm{mfp}^{-1}
	    &\sim \frac{1}{2E_1} \sum_{12\to34} \int \prod_{i=2,3,4}\frac{\dd[3]p_i}{(2\pi)^3 2 E_i} (2\pi)^4 \delta^4 \qty(p_1^\mu + p_2^\mu - p_3^\mu - p_4^\mu) \abs{\mathcal M_{12 \to 34}}^2 f_2 (1 \mp f_3) (1 \mp f_4)\\
	    & \sim \sum_{12\to34} n_2  \qty(\sigma v_\mathrm{rel})_{12\to34}.
	\end{split}\label{mfp}
	\end{equation}
It is noteworthy that the mean free path of a particle labelled as $1$ is determined by the cross sections and incoming particles densities labelled as $n_2$ involved in the reactions. To estimate the mean free path given in Eq.(\ref{mfp}), it should be pointed that the sum runs over all possible $ 12 \to 34 $ processes, and the number of processes involved in the left-right symmetric model are typically of order $ \order{1 \sim 10} $. The scattering number density $n_2$ can be estimated as $ n_2 \lesssim \zeta_3 \, T^3 / 4 \pi^3 \sim 0.01 \, T^3 $ and the total cross section can be computed as $ \qty(\sigma v_\mathrm{rel})_{12\to34} \sim \lambda^2 \, T^{-2} / 16 \pi \sim 0.02 \, \lambda^2 \, T^{-2} $ where typically, the couplings $ \lambda $ can be taken in the perturbation range $ \order{0.1 \sim 1} $. Thus we can obtain the free mean path as $ \lambda_\mathrm{mfp}^{-1} $ to be $ 10^{-5} \sim 10^{-3} \, T $ or so. The validity of free particle approximation and LTE approximation is discussed in the following subsections.

\subsection{Free particle approximation}
	\label{subsec:freeparticle}

In the case where the mean free path $\lambda_\mathrm{mfp}$ is much larger than the bubble wall thickness $L_w$
\begin{equation}
		\label{eq:Lwsmall}
		\lambda_\mathrm{mfp} \gg \gamma_w^{-1} L_w,
\end{equation}
we can neglect $ C[f] $ in Eq. \eqref{eq:Boltzmann}. It is also found that when the condition Eq. (\ref{eq:Lwsmall}) holds, the profile of bubble does not change the distributions of particles. Thus we leave the validation of Eq. \eqref{eq:Lwsmall} after the $ v_w $ have been done where in \ref{subsec:vw}.
	
The interaction between particles and scalar field can be more easily studied in the bubble wall frame. In such a frame, the conservation laws of energy-momentum tensor can be expressed as 
	\begin{equation}
		\label{eq:conserveWallframe}
		\pd_z T^{zz} = 0,\quad
		\pd_z T^{z0} = 0.
	\end{equation}
	Inserting Eq. \eqref{eq:TmunuField} and \eqref{eq:TmunuPlasma} into \eqref{eq:conserveWallframe}, we arrive at the equation of motion for scalar field, which has the following form 
	\begin{equation}
		\label{eq:scalarEoM}
		0 = \pdv{\phi}{z} \qty(\pdv[2]{\phi}{z} - \pdv{V_\mathrm{vac}}{\phi} - \pdv{\phi} \sum_i \int \frac{p_z^2 \dd[3]p}{(2 \pi)^3 E_i} f_i(\vb p, \vb x)).
	\end{equation}
Meanwhile, the $t$ dependency of distribution function can also be eliminated and the Boltzmann equations can be modified as 
	\begin{equation}
		\label{eq:BoltzmannWallFrame}
		\qty(\frac{p_z}{E} \pdv{z} - \frac{m}{E} \pdv{m}{z} \pdv{p_z}) f(p_z, p_\perp, z) = 0.
	\end{equation}
For a single bubble wall, the mass parameter of a particle $ m $ decreases monotonically along $ z $-direction, thus we can use the following substitution
	\begin{equation}
		\pdv{z} = \pdv{m}{z} \pdv{m}.
	\end{equation}
	This leads a general solution for Eq. \eqref{eq:BoltzmannWallFrame}
	\begin{equation}
		\label{eq:generalSolution}
		f = f(p_z^2 + m^2, p_\perp, z).
	\end{equation}
	By noting the mass-energy relation $ E = \sqrt{p_z^2 + p_\perp^2 + m^2} $, it should be pointed out that the solution given in Eq. \eqref{eq:generalSolution} has taken into account the energy conservation for every single particle species.
	\par
\begin{figure}[h!]
		\centering
    \includegraphics[width = 0.7\columnwidth]{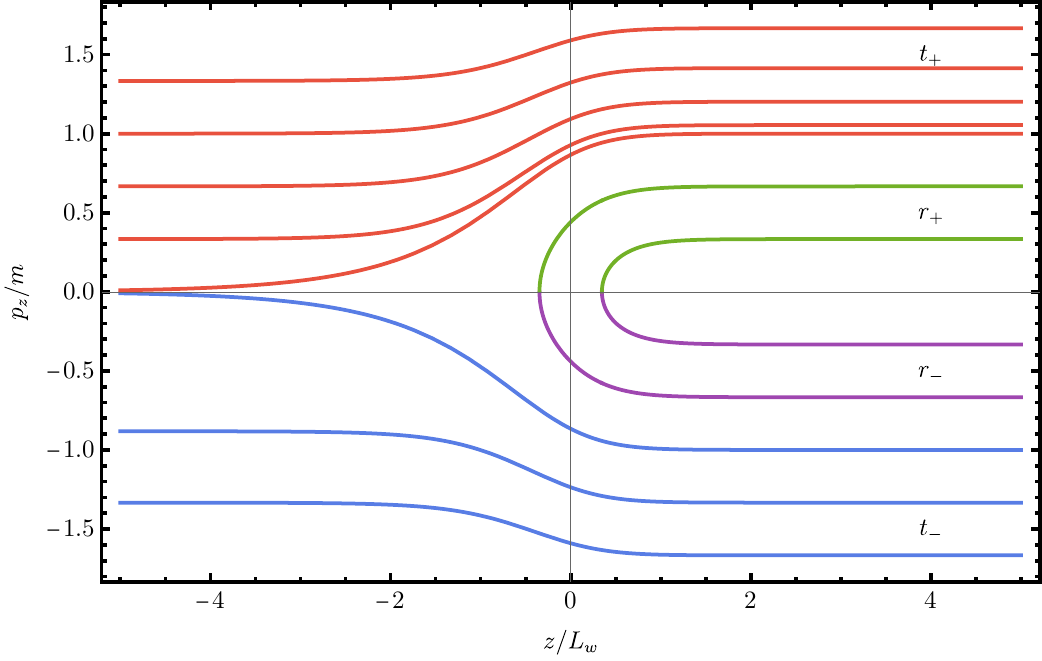}
		\caption{Trajectories of particles from different branches in phase space given in Eqs. \eqref{eq:ft-}, \eqref{eq:fr} and \eqref{eq:ft+}, are shown, where we set $m_s=0$. Sign in subscript labels the moving direction of a particle and the reflection branch has been split into two parts.}
		\label{fig:zpz}
	\end{figure}

There are two boundaries, which correspond to $ z \to - \infty$ where the plasma is in the broken phase and $ z \to + \infty$ where the plasma is in the symmetric phase. Near the regions of these two boundaries, the distribution functions of a particle that moving toward the wall can be approximately described by the distribution functions of equilibrium. Since we are working in the bubble wall frame, the distribution function is boosted in $- z$-direction, i.e.
	\begin{equation}
		\begin{split}
			f(p_z) &= \frac{1}{\exp(\beta_s \gamma_w \qty(E(m_s) + v_w p_z)) \pm 1},\quad
			p_z < 0, z \to + \infty,\\
			f(p_z) &= \frac{1}{\exp(\beta_b \gamma_w \qty(E(m_b) + v_w p_z)) \pm 1},\quad
			p_z > 0, z \to - \infty,
		\end{split}
	\end{equation}
where the subscripts $ s $ and $ b $ denote symmetric and broken phases where the particles are coming from, but $ \beta_{s, b}^{-1} $ are not necessarily equal to the temperature of the symmetric or broken phase. We will address this issue in the discussion on the results shown by FIG. \ref{fig:tttt}. Applying these boundary conditions to Eq. \eqref{eq:generalSolution}, the solution can be written in three branches 
    \cite{cite:FieldTh21}: 
	\begin{enumerate}
		\item[B1:] transmission from broken phase $ t_+ $
		\begin{equation}
			\label{eq:ft-}
			f (p_z, p_\perp,z) = \frac{1}{\exp(\beta_b \gamma_w \qty(E + v_w p_{z,b})) \pm 1},\quad
			p_z \ge + \sqrt{m_b^2 - m^2(z)};
		\end{equation}
		\item[B2:] reflection $ r $
		\begin{equation}
			\label{eq:fr}
			f (p_z, p_\perp,z) = \frac{1}{\exp(\beta_s \gamma_w \qty(E + v_w p_{z,s})) \pm 1},\quad
			|p_z| < \sqrt{m_b^2 - m^2(z)};
		\end{equation}
		\item[B3:] transmission from symmetric phase $ t_- $
		\begin{equation}
			\label{eq:ft+}
			f (p_z, p_\perp,z) = \frac{1}{\exp(\beta_s \gamma_w \qty(E + v_w p_{z,s})) \pm 1},\quad
			p_z \le - \sqrt{m_b^2 - m^2(z)},
		\end{equation}
	\end{enumerate}
	where $ p_{z,s} \equiv \sqrt{p_z^2 + m^2(z) - m_s^2} $, similar to $ p_{z,b} $. A schematic picture of these three branches are shown in \figurename{} \ref{fig:zpz}, where $z \to - \infty$ is the broken phase, and $z \to \infty$ corresponds to the symmetry phase.
	
	It should be pointed out that for reflection branch, the solution given in Eq. (\ref{eq:fr}) in this work is different from that given in \cite{cite:FieldTh21}, i.e. a correct result has no $ \sign[p_z] $ in front of $ v_w $. If the $ \sign[p_z] $ is added in front of $ v_w $, it is found that the energy conservation might be broken. 
	
	In order to determine the bubble wall velocity, we need use the energy-momentum tensors of plasma, which can be computed when the distribution functions of particles in the plasma are known. Since we have neglected the collisions between particles, the energy flux should be constant along $z$-direction for every branch separately. Thus we should have
	\begin{equation}
		T_{t_\pm}^{0z}(z \to + \infty) = T_{t_\pm}^{0z}(z \to - \infty),\qq{and}
		T_{r_+}^{0z}(z \to +\infty) = - T_{r_-}^{0z}(z \to +\infty),
	\end{equation}
	where $ T_{t, r}^{0z} $ can be calculated by using Eq. \eqref{eq:TmunuPlasma}, which contribute to $ T^{\mu\nu}_\mathrm{plasma} $ by definition. 
	While it should be mentioned that $ T_{r_+}^{0z} $ ($ T_{r_-}^{0z} $) is defined as being integrated for momentum modes $ p_z < 0 $ ($ p_z > 0 $) only.
	
	By using $ E \dd{E} = p_\perp \dd{p_\perp} $, we can integrate out $ p_\perp $ first. Thus only the integration over $p_z$ is left. For $ T_{t_\pm}^{0z} $, we arrive at the following results 
	\begin{equation}
		\begin{split}
			T_{t_+}^{0z} &= + \int^\infty_0 \frac{p_z \dd{p_z}}{(2\pi)^2} \qty(
			\pm \frac{E_{z, b}}{\beta_b \gamma_w} \ln(1 \pm \exp(- \beta_b \gamma_w \qty(E_{z, b} + v_w p_z)))
			\mp \frac{1}{\beta_b^2 \gamma_w^2} \Li_2(\exp(- \beta_b \gamma_w \qty(E_{z, b} + v_w p_z)))),\\
			T_{t_-}^{0z} &= - \int^\infty_{\sqrt{\Delta m^2}} \frac{p_z \dd{p_z}}{(2\pi)^2} \qty(
			\pm \frac{E_{z, s}}{\beta_s \gamma_w} \ln(1 \pm \exp(- \beta_s \gamma_w \qty(E_{z, s} - v_w p_z)))
			\mp \frac{1}{\beta_s^2 \gamma_w^2} \Li_2(\exp(- \beta_s \gamma_w \qty(E_{z, s} - v_w p_z)))),
		\end{split}
	\end{equation}
	where $ \Delta m^2 = m_b^2 - m_s^2 $, $ E_{z, s} = \sqrt{p_z^2 + m_s^2} $, and $ E_{z, b} = \sqrt{p_z^2 + m_b^2}$. 
	
To derive $ T^{zz}(z) $, it is noticed that the reflection branch does contribute which is different from $T^{0z}(z)$. After performing integration over $p_\perp$, we arrive the following results, which can be organized into three branches,
	\begin{enumerate}
		\item[B1:] transmission from broken phase $ t_+ $
		\begin{equation}
			\label{eq:Tt+}
			T^{zz}_{t_+}(z) = \pm \frac{1}{\beta_b \gamma_w} \int^\infty_{\sqrt{m_b^2 - m^2(z)}} \frac{p_z^2 \dd{p_z}}{(2 \pi)^2} \ln(1 \pm \exp(-\beta_b \gamma_w \qty(E_z + v_w p_{z,b})));
		\end{equation}
		\item[B2:] reflection $ r $
		\begin{equation}
			\label{eq:Tr}
			T^{zz}_{r}(z) = \pm \frac{2}{\beta_s \gamma_w} \int^{\sqrt{m_b^2 - m^2(z)}}_0 \frac{p_z^2 \dd{p_z}}{(2 \pi)^2} \ln(1 \pm \exp(-\beta_s \gamma_w \qty(E_z - v_w p_{z,s})));
		\end{equation}
		\item[B3:] transmission from symmetric phase $ t_- $
		\begin{equation}
			\label{eq:Tt-}
			T^{zz}_{t_-}(z) = \pm \frac{1}{\beta_s \gamma_w} \int^\infty_{\sqrt{m_b^2 - m^2(z)}} \frac{p_z^2 \dd{p_z}}{(2 \pi)^2} \ln(1 \pm \exp(-\beta_s \gamma_w \qty(E_z - v_w p_{z,s}))),
		\end{equation}
	\end{enumerate}
	where $ E_z \equiv{p_z^2 + m^2(z)} $. It is observed that in the limit $ v_w \to 0 $, $ \sum_{t_\pm,r} T^{zz}_{t_\pm,r}(\beta = 1 / T, m(\phi), v_w) = V_1^{T \neq 0}(T, \phi) $ holds exactly. 
	
To calculate the quantities $T^{0z}$ and $ T^{zz} $, we should sum over all particle species in the plasma. Here, for the sake of simplicity, we have dropped the symbol of sum. 
		
When the bubble wall moves steadily in the plasma, the pressure from the inside of bubble wall and the pressure from the outside of the bubble wall should be balanced. Thus in terms of energy conservation laws, the following relations should hold 
	\begin{equation}
		\label{eq:DTzzeq}
		- \Delta T^{zz}_\mathrm{field} = \Delta T^{zz}_\mathrm{plasma} \equiv \Delta T^{zz}_{i_+} + \Delta T^{zz}_{i_-},
	\end{equation}
	where the quantities in Eq. (\ref{eq:DTzzeq}) are defined as
	\begin{equation}
		\begin{cases}
			\Delta T^{zz}_\mathrm{field} \equiv - V_\mathrm{vac}(z \to +\infty) + V_\mathrm{vac}(z \to -\infty);\\
			\Delta T^{zz}_{i_+} \equiv T^{zz}_{t_+}(+\infty) - T^{zz}_{t_+}(-\infty),
			& \text{particles incident from broken phase;}\\
			\Delta T^{zz}_{i_-} \equiv T^{zz}_{t_-}(+\infty) - T^{zz}_{t_-}(-\infty) + T^{zz}_{r}(+\infty),
			& \text{particles incident from symmetric phase}.
		\end{cases}
	\end{equation}
	\begin{figure}[h!]
	\centering
    \includegraphics[width = 0.9\columnwidth]{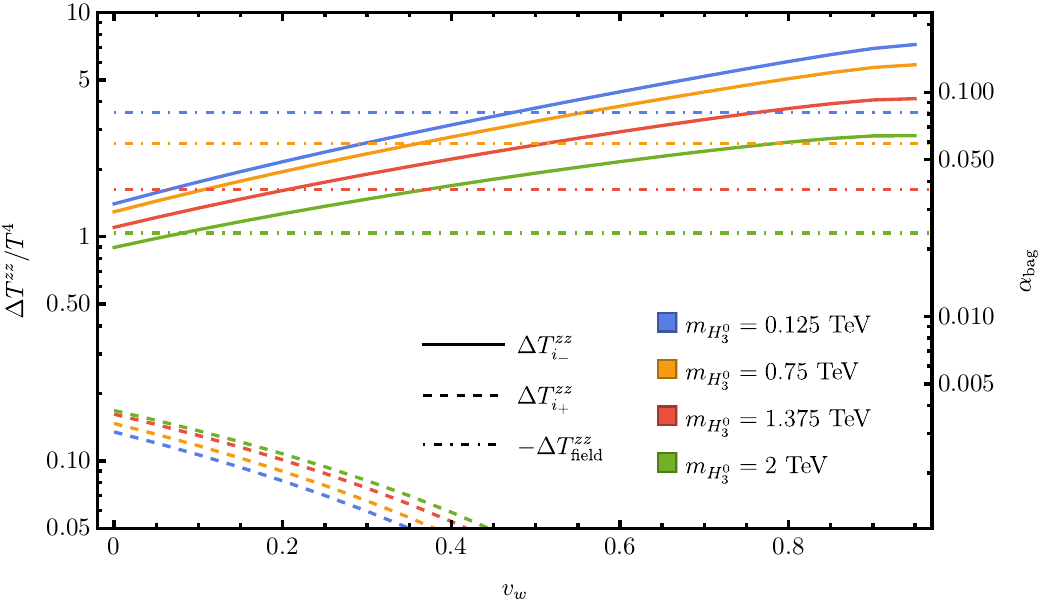}
		\caption{Model parameter and $v_w$ dependence of $T^{zz}$ is demonstrated. The $ \Delta T^{zz}_{i_\pm} $ for different $ m_{H_3^0} $ are represented by solid (dashed) lines, with $ m_{H_1^0} = m_{H_2^0} = 10 $ TeV and $ m_{H_2^{++}} = 8 $ TeV. The $ \Delta T^{zz}_\mathrm{field}$ for different case are represented by horizontal dash-dotted lines.}
		\label{fig:vwTzz}
	\end{figure}
	
In \figurename{} \ref{fig:vwTzz}, we demonstrate the features of $ \Delta T^{zz}$ for $ m_s = 0 $ with different bubble wall velocity $ v_w $ and $ \sqrt{\Delta m^2} / T $ values. The y-axis denotes $\Delta T^{zz}$ and the x-axis denotes the bubble wall velocity $v_w$. In the plot, the vacuum pressure difference $ \Delta T^{zz}_\mathrm{field} $ is represented by the horizontal dash-dotted lines. It is observed that $ \Delta T^{zz}_{i_-}/T^4 $ increases with the increase of $v_w$ while $ \Delta T^{zz}_{i_+}/T^4 $ decreases with the increase of $v_w$. Due to the fact that the term $ \Delta T^{zz}_{i_-} $ contributes dominantly to the quantity $ \Delta T^{zz}_\mathrm{plasma} $ and $ \Delta T^{zz}_\mathrm{field} $ does not change with $ v_w $, roughly speaking, one can read off $ v_w $ by eyes which should be at the place where  $ \Delta T^{zz}_{i_-} \approx - \Delta T^{zz}_\mathrm{field} $. Obviously, to find the exact result for $ v_w $, one need to solve the Eq. \eqref{eq:DTzzeq}
    \begin{equation}
    \label{eq:vwDTzz}
		0 = \Delta T^{zz}_\mathrm{field} \qty(\phi_*) + \Delta T^{zz}_{i_+} \qty(\phi_*, \beta_b, v_w) + \Delta T^{zz}_{i_-} \qty(\phi_*, \beta_s, v_w)
		\equiv \Delta T^{zz}(\phi_*, \beta_s, \beta_b, v_w),
    \end{equation}
    where $ \phi_* $ is the VEV in broken phase and is determined by the following condition 
    \begin{equation}
    \label{eq:phiDTzz}
        0 = \eval{\pdv{\Delta T^{zz}(\phi, \beta_s, \beta_b, v_w)}{\phi}}_{\phi = \phi_*}.
    \end{equation}
There are still two independents $ \beta_{b,s} $ parameters in the equation remains undetermined. Generally speaking, one can use $ T_N^{-1} $ ($T_N$ is the nucleation termparature) as an approximate solution. However, in this work, we make use of hydrodynamics which has been discussed in \ref{subsec:LTE}. This helps us to determine $ \beta_{b,s} $ by matching $ T^{zz} $ and $ T^{0z} $ calculated from two different approaches.

It is interesting to compare $\Delta T^{zz}_\mathrm{field}$ with the parameter $ \alpha_{bag} $ in the bag-model \cite{cite:ErgBdg}. Since $ - \Delta T^{zz}_\mathrm{field} $ is proportional to bag constant by the following relation, 
	\begin{equation}
		\frac{\Delta T^{zz}_\mathrm{field}}{T^4} = \frac{\Delta V_\mathrm{vac}}{T^4} \sim \alpha_\mathrm{bag} \frac{g_* \pi^2}{30},\qq{where}
		g_* = 134.
	\end{equation}
thus we can solve out $\alpha_{bag}$. The values of $\alpha_{bag}$ is depicted by the right y-axis in \figurename{} \ref{fig:vwTzz}, which is in the range of $0.01 \sim 0.1$ when model parameters are taken in the range specified.

	
\subsection{Local thermal equilibrium approximation}
\label{subsec:LTE}
To calculate the terminal bubble wall velocity, we can assume that the bubble radius $ r_w $ is close to the mean bubble separation $ R_* $ \cite{cite:Hindmarsh19}
	\begin{equation}
		r_w \sim R_*, \qq{where}
		R_* H_* = (8 \pi)^{1 / 3} v_w (H_* / \beta).
	\end{equation}
For the typical value of $ \beta / H_* $ shown in \figurename{} \ref{fig:alphabeta} and the nucleation temperature $ T_N \simeq 2.7\, \mathrm{TeV} $, we have $ R_* \sim 10^{11} \, T_N $. In such a case, the bubble radius $ r_w $ can be sufficiently large, such that $ r_w \gg \lambda_\mathrm{mfp} $, the particles far from the bubble wall can be treated as in LTE.

In the regions far from the bubble wall, the field-plasma system can be described by perfect fluid
	\begin{equation}
		T^{\mu\nu}_\mathrm{LTE} = w u^\mu u^\nu - p g^{\mu\nu},
	\end{equation}
where the pressure $ p $, energy density $ e $ and enthalpy $ w $ are given by
	\begin{equation}
		\label{eq:defthermalfunc}
		p = - V_\mathrm{eff}(T, \phi_*(T)),\quad
		e = T \pdv{p}{T} - p = V_\mathrm{eff} - T \pdv{V_\mathrm{eff}}{T},\quad
		w = e + p = - T \pdv{V_\mathrm{eff}}{T}.
	\end{equation}
In the symmetric phase, we take $ \phi_*(T) = 0 $. While in broken phase, we take
	\begin{equation}
		0 = \eval{\pdv{V_\mathrm{eff}(T, \phi)}{\phi}}_{\phi = \phi_*(T)}
	\end{equation}
	as the definition of $ \phi_* $.
	The fluid 4-velocity $ u^\mu $ is parameterized as
	\begin{equation}
		u^\mu = \gamma (1, \vb{v}), \qq{where} \gamma = 1 / \sqrt{1 - v^2},
	\end{equation}
where $ \vb v $ is 3-velocity. The conservation law of energy-momentum can then be projected on $ u^\mu $ and $ \bar{u}^\mu \equiv \gamma (v, \vb{v} / v) $. 
	\begin{equation}
		\label{eq:hydroConserve}
		u_\nu \pd_\mu T^{\mu\nu}_\mathrm{LTE} = \pd_\mu (w u^\mu) - u^\mu \pd_\mu p = 0,\quad
		\bar{u}_\nu \pd_\mu T^{\mu\nu}_\mathrm{LTE} = w u^\mu \bar{u}^\nu \pd_\mu u_\nu - \bar{u}^\mu \pd_\mu p = 0
	\end{equation}

The conservation law of energy-momentum holds everywhere in the plasma, we can apply it to find the profiles of sound shell. Since there is no specific length scale in Eq. (\ref{eq:hydroConserve}), we assume that the solution of Eq. \eqref{eq:hydroConserve} for the profile of sound shell is spherical and self-similar, where $ p $, $ e $ and $ v $ are functions of $ \xi \equiv r / t $. Here $r$ denotes the radius of bubble and t is the time elapse after the bubble is formed. Thus, in reality, the parameter $\xi$ can only take the value in the range $0 \leq \xi \leq 1$. Then Eq. (\ref{eq:hydroConserve}) can take the following form 
	\begin{equation}
		\label{eq:epvHydroEq}
		\begin{split}
			(\xi - v) \frac{\pd_\xi e}{w} &= 2 \frac{v}{\xi} + [1 - \gamma^2 v (\xi - v)] \pd_\xi v, \\
			(1 - \xi v) \frac{\pd_\xi p}{w} &= \gamma^2 (\xi - v) \pd_\xi v.
		\end{split}
	\end{equation}
	With the definition of speed of sound $ c_s^2 = (\pd p / \pd T) / (\pd e / \pd T) $, Eq. \eqref{eq:epvHydroEq} can be written as
	\begin{equation}
		\label{eq:vtHydroEq}
		\begin{split}
			2 \frac{v}{\xi} &= \qty(\frac{\mu^2}{c_s^2} - 1) \gamma^2 (1 - \xi v) \pd_\xi v, \\
			2 \frac{v}{\xi} &= \qty(\frac{\mu^2}{c_s^2} - 1) \frac{(1 - \xi v)^2}{\xi - v} \frac{\pd_\xi T}{T},
		\end{split}
	\end{equation}
	where $ \mu(\xi, v) \equiv (\xi - v) / (1 - \xi v) $ is the fluid velocity in local frame.
	
When considering the temperature dependence of $ c_s $, the second line in Eq. \eqref{eq:vtHydroEq} should also be solved simultaneously. Due to the singularities appear in Eq. \eqref{eq:vtHydroEq} at $ \mu(\xi, v) = c_s $, it is convenient to introduce a parameter $ \tau $ to rewrite the equation into a parametric form \cite{cite:ErgBdg, cite:Hindmarsh19}
\begin{equation}
\label{eq:vtHydroEqParam}
\begin{split}
\pd_\tau \xi &= \xi [(\xi - v)^2 - c_s^2 (1 - \xi v)^2],\\
\pd_\tau v &=2 v c_s^2 (1 - v^2) (1 - \xi v),\\
\frac{\pd_\tau T}{T} &= 2 v c_s^2 (\xi -v).
\end{split}
\end{equation}
	
Since we are working in the limit of $ R \gg L_w $, the bubble wall can be treated as a step function. This will induce discontinuities in $ p $, $ e $ and $ v $. After boosting into the bubble wall frame, the conservation law become
\begin{equation}
\label{eq:tzzt0zhydro}
\begin{split}
T^{0r}_\mathrm{LTE} &= w_+ \gamma_+^2 v_+ = w_- \gamma_-^2 v_-,\\
T^{rr}_\mathrm{LTE} &= w_+ \gamma_+^2 v_+^2 + p_+ = w_- \gamma_-^2 v_-^2 + p_-,
\end{split}
\end{equation}
where $ + $ stands for quantities in symmetric phase and $ - $ for broken phase. By introducing $ \gamma \equiv \cosh y $ and $ \gamma v \equiv \sinh y $, Eq. \eqref{eq:tzzt0zhydro} can be cast into
\begin{equation}
\label{eq:vbvstbts}
\frac{v_+}{v_-} = \frac{e_- + p_+}{e_+ + p_-},\quad
v_+ v_- = \frac{p_+ - p_-}{e_+ - e_-},
\end{equation}
where $ p_\pm $ and $ e_\pm $ are functions of $ T_\pm $.
\begin{figure}[h!]
\centering
\includegraphics[width = 1\columnwidth]{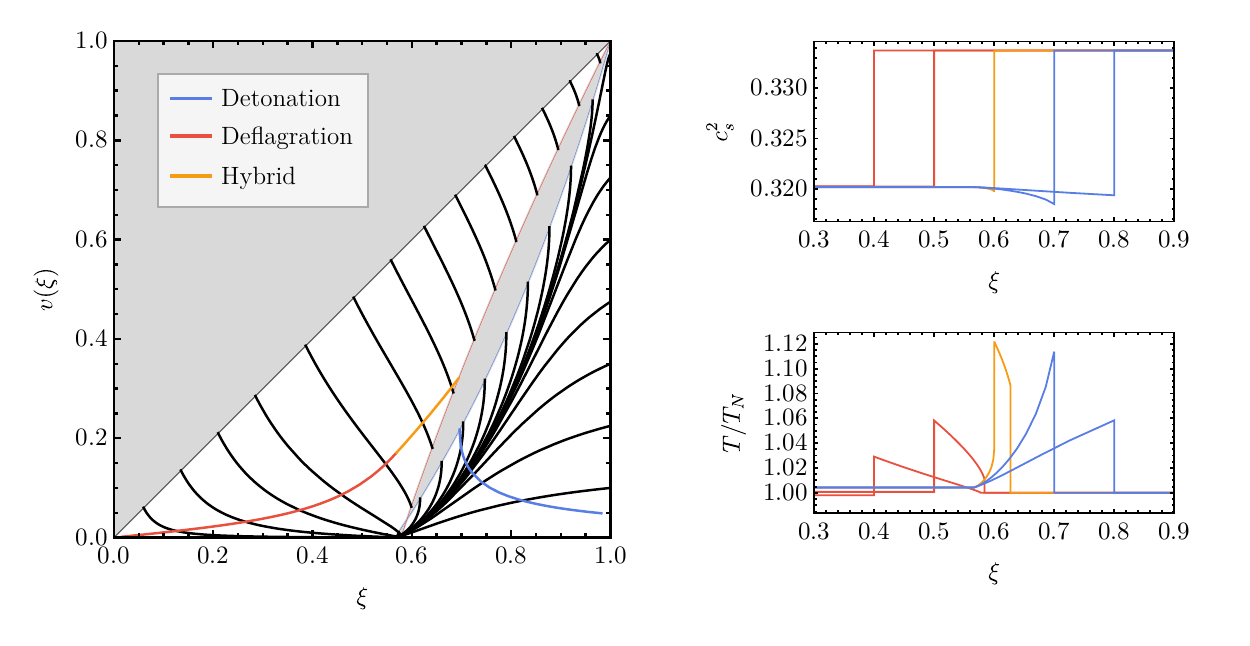}
\caption{Left panel: Black lines are solutions of Eq. \eqref{eq:vtHydroEqParam} for $ v(\xi) $. Solutions for Eq. \eqref{eq:detbnd}, \eqref{eq:defbnd} and \eqref{eq:hybbnd} are shown in the blue, red and orange lines, respectively. Shaded regions are physically forbidden. 
Right panel: profiles for $ c_s^2 $ and $ T / T_N $ as functions of $ \xi $ for $ \xi_w = 0.4 \sim 0.8 $. The blue, red and orange lines are solutions of Eq. \eqref{eq:vbvstbts} for detonation, deflagration and hybrid.}
\label{fig:xiv}
\end{figure}
There are total four unknowns in Eq. \eqref{eq:vbvstbts}, which can be taken as $ v_\pm $ and $ T_\pm $. Two of them must be input to make the equations solvable. The motion of the fluid has been divided into three modes\cite{cite:ErgBdg}, i.e. deflagration, hybrid, detonation.

{\bf Detonation mode} happens when $ v_J \le v_w $, where the Jouguet velocity $ v_J $ is determined by finding the minimal $ v_+ $ on $ v_+ > v_- $ branch that satisfies Eq. \eqref{eq:vbvstbts}
	\begin{equation}
		v_J = \underset{v_-}{\min} \ v_+(T_+ = T_N, v_-),\qq{with}
		v_+ > v_-.
	\end{equation}
In this mode, the bubble wall is follow by rarefaction wave, while the fluid in front of the bubble wall remains in steady. Thus we have following relations
	\begin{equation}
	\label{eq:detbnd}
		T_+ = T_N,\quad
		v(\xi_w) = \mu(\xi_w, v_-),\qq{where}
		\xi_w = v_+.
	\end{equation}
	
{\bf Deflagration mode} happens when $ v_w \le c_{s,-} $. In this mode, the fluid behind the bubble wall remains steady, while a shock front can appear in front of the bubble wall. Thus we have
	\begin{equation}
	\label{eq:defbnd}
		T_\mathrm{sh, +} = T_N,\quad
		v(\xi_w) = \mu(\xi_w, v_+),\qq{where}
		\xi_w = v_-.
	\end{equation}
	$ T_\mathrm{sh, +} $ is the temperature in front of the shock. The location of chock front $ \xi_\mathrm{sh} $ can also be determined by Eq. \eqref{eq:vbvstbts} with relation
	\begin{equation}
		v(\xi_\mathrm{sh}) = \mu(\xi_\mathrm{sh}, v_\mathrm{sh, -}),\qq{where}
		\xi_\mathrm{sh} = v_\mathrm{sh, +}.
	\end{equation}
	The shock front in deflagration mode can vanish for a rather small $ v_w $, in such case $ v(\xi) $ falls to 0 continually at $ \xi = c_{s, +} $.
	
{\bf Hybrid mode} happens when $ c_{s, -} < v_w < v_J $. In this mode, the fluid both behind and in front of the bubble wall has been perturbed. In bubble wall frame, the rarefaction wave begins with fluid velocity $ v_- = c_{s, -} $. So we have relations
	\begin{equation}
	\label{eq:hybbnd}
		v_\mathrm{det}(\xi_w) = \mu(\xi_w, c_{s, -}),\qq{and}
		v_\mathrm{def}(\xi_w) = \mu(\xi_w, v_+),
	\end{equation}
	where $ v_\mathrm{det} $ and $ v_\mathrm{def} $ refer to fluid velocity behind and in front of the wall respectively. 

On the left panel of \figurename{} \ref{fig:xiv}, we show the solutions of Eq. \eqref{eq:vtHydroEq} which are depicted in black lines, where shaded regions are physically forbidden. Three modes are shown by blue, red, and orange lines, respectively. On the right panel of \figurename{} \ref{fig:xiv}, we show the profiles of $ c_s^2 $ and $ T / T_N $. It should be pointed out that for different modes, the relation between $ v_\pm $ and $ T_\pm $ versus $ v_w $ and $ T_N $ are different.

\subsection{Bubble wall velocity and thickness}
\label{subsec:vw}
\begin{figure}[h!]
\centering
\includegraphics[width = 0.7\columnwidth]{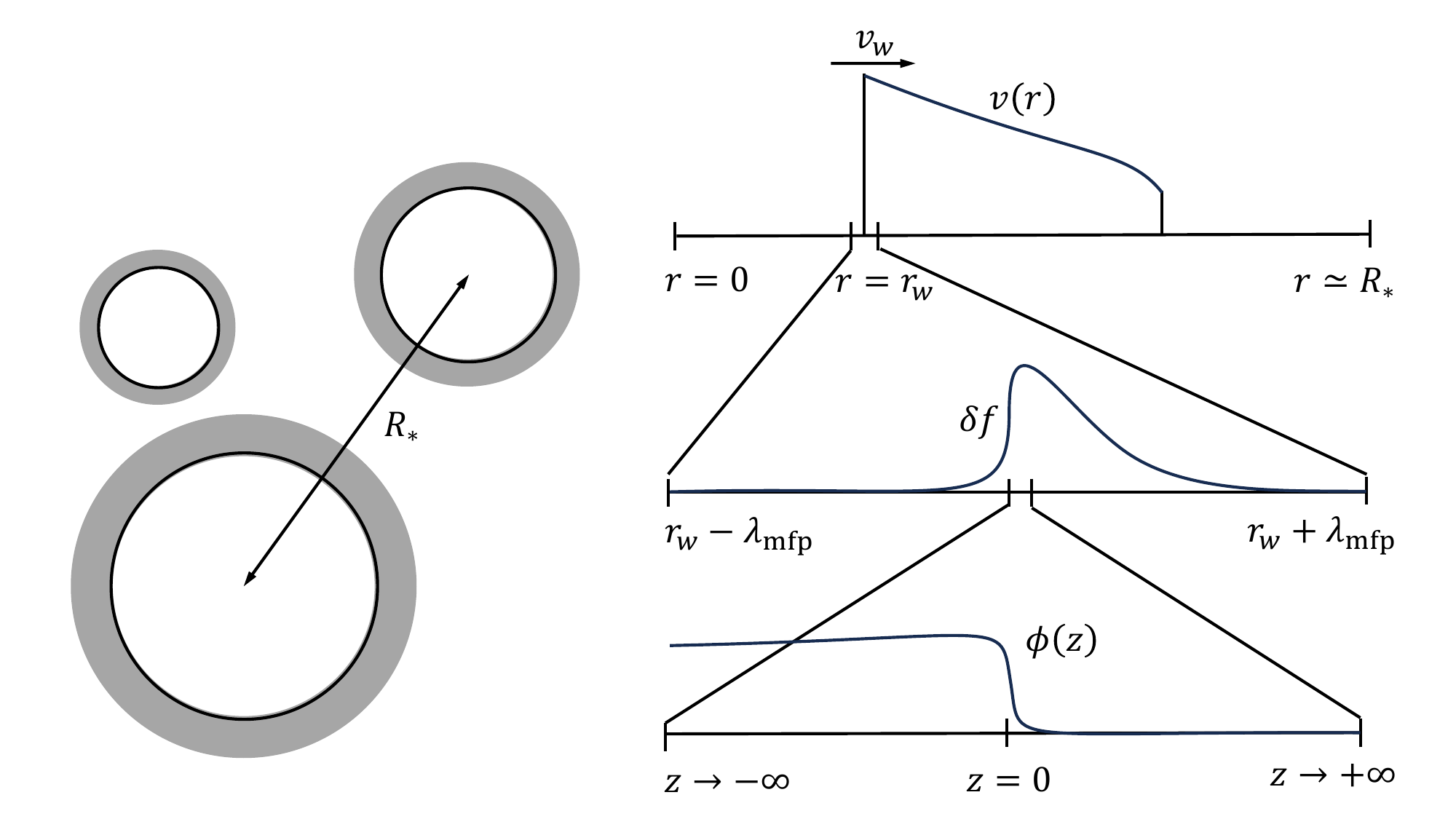}
\caption{Different approximation works at different length scale, while the EMT conservation holds everywhere. When $ v_w $ approaches its terminal value, the system becomes steady, we assume the profile of $ \delta f $ won't change with time any more. Thus energy(momentum) flux should be equal at $ z = \pm \infty $ and $ r = r_w $.}
\label{fig:scale}
\end{figure}

We have introduced two different approaches to describe the field-plasma system, the free particle approximation and the local thermal equilibrium, and they work at different length scales.
These two approaches should compatible with each other. When the bubble radius $ r_w $ grows to $ r_w \simeq R_* $, if the wall does not runaway, the field-plasma system becomes steady and the wall velocity approaches to its terminal value $ v_w $. At this stage, the profile of scalar field is steady in wall frame, while the profile of fluid variables, i.e. $ v $ and $ T $, broaden in $ r $-direction with time as a self-similar solution. \figurename{} \ref{fig:scale} gives a schematic picture on what happens at this stage for different length scales. Within region where $ r = r_w \pm \lambda_\mathrm{mfp} $, the fluid profile appears to be discontinuities, and the non-equilibrium effect become significant. The incident, reflected and transmitted particles collide with each other, which makes the deviation $ \delta f \equiv f - f_{eq} $ dissipates in this region. Out of this region, $ \delta f $ vanishes, bring the system back to equilibrium, where LTE approximation works. If we further look into a narrower region around bubble wall, i.e. in $ L_w $ length scale, the particle collision becomes less important, while the variation in $ \phi $ dominates the motion of particles. In other word, we should map the coordinate $ z \in (-\infty, +\infty) $ into $ r \in (r_w - \lambda_w, r_w + \lambda_w) $, with $ \lambda_w \gtrsim L_w $.

Despite these different approaches work at different length scales, the conservation of energy and momentum always holds. And since the system is steady, the profile of $ \phi $ and $ \delta f $ should not vary with time. In other words, in the bubble wall frame, the energy and momentum accumulated in $ |r - r_w| < \lambda_\mathrm{mpf} $ region are constant. While the energy and momentum flux at $r = r_w \pm \lambda_w $ should be equal, and their value should match the flux at the discontinuity in LTE approximation. Thus we come up with a method described in the following part of this subsection.

For the convenience, we define the following quantities
\begin{equation}
\label{eq:tzzt0zeqdef}
\begin{split}
\begin{cases}
T^{0z}_{\phi = 0}
				\equiv \qty(T^{zz}_{t_+} + T^{0z}_{t_-})_{\phi = 0} \\
				T^{0z}_{\phi = \phi_*} \equiv \qty(T^{zz}_{t_+} + T^{0z}_{t_-})_{\phi = \phi_*}
				\\
				T^{zz}_{\phi = 0}
				\equiv \qty(T^{zz}_{t_+} + 2 T^{zz}_{r} + T^{zz}_{t_-} - V_\mathrm{vac})_{\phi = 0}\\
				T^{zz}_{\phi = \phi_*} \equiv \qty(T^{zz}_{t_+} + T^{zz}_{t_-} - V_\mathrm{vac})_{\phi = \phi_*}
			\end{cases}
		\end{split}
	\end{equation}
where the subscript $\phi=0$ denotes the symmetric phase and the subscript $\phi=\phi_*$ denotes the broken phase. Thus the requirement of energy and momentum flux should be equal on both sides of bubble wall leads to the following identities
	\begin{equation}
		\label{eq:tzzt0zeq}
		\begin{split}
			\begin{cases}
				T^{0z}_{\phi = 0}
				= T^{0z}_{\phi = \phi_*}\\
				T^{zz}_{\phi = 0}
				=T^{zz}_{\phi = \phi_*}
			\end{cases}
		\end{split}
	\end{equation}
where $ \pd T^{zz}_{\phi = \phi_*} / \pd \phi = 0$ has been used. The first line of Eq. \eqref{eq:tzzt0zeq} is the energy flux equation, which always holds automatically due to energy conservation. However, the second line of Eq. \eqref{eq:tzzt0zeq} is the momentum flux equation, which does not always hold. The reason lies in the fact that the translation invariant is broken by the wall. Thus we need a set of free parameters $ (v_w, \beta_b, \beta_s) $ to calculate $ T^{0z} $ and $ T^{zz} $ and the second line of Eq. \eqref{eq:tzzt0zeq} can help to determine one of these three free parameters. 
	
In local thermal equilibrium approximation, we also have components of energy-momentum tensors $ T^{0r}_\mathrm{LTE} $ and $ T^{rr}_\mathrm{LTE} $, which only take $ v_w $ as free parameter.  It is possible to determine all these three free parameters $ (v_w, \beta_b, \beta_s) $ by using Eq. (\ref{eq:tzzt0zeq}) and by further requiring
	\begin{equation}
		\label{eq:tzztrreq}
		T^{0z} \simeq T^{0r}_\mathrm{LTE}\qq{and}
		T^{zz}_{\phi = 0} = T^{zz}_{\phi = \phi_*} \simeq T^{rr}_\mathrm{LTE}.
	\end{equation}
To identify $T^{0z}$ ($T^{zz}$) and $T^{0r}_\mathrm{LTE}$ ($T^{rr}_\mathrm{LTE}$) can only make sense when the radius of bubble wall is large enough and the movement of bubble wave can be approximated by two parallel planes separated by a distance $L_w$, which can be true when the bubble wall velocity is steady. 

	\begin{figure}[h!]
		\centering
		\includegraphics[width = 1\columnwidth]{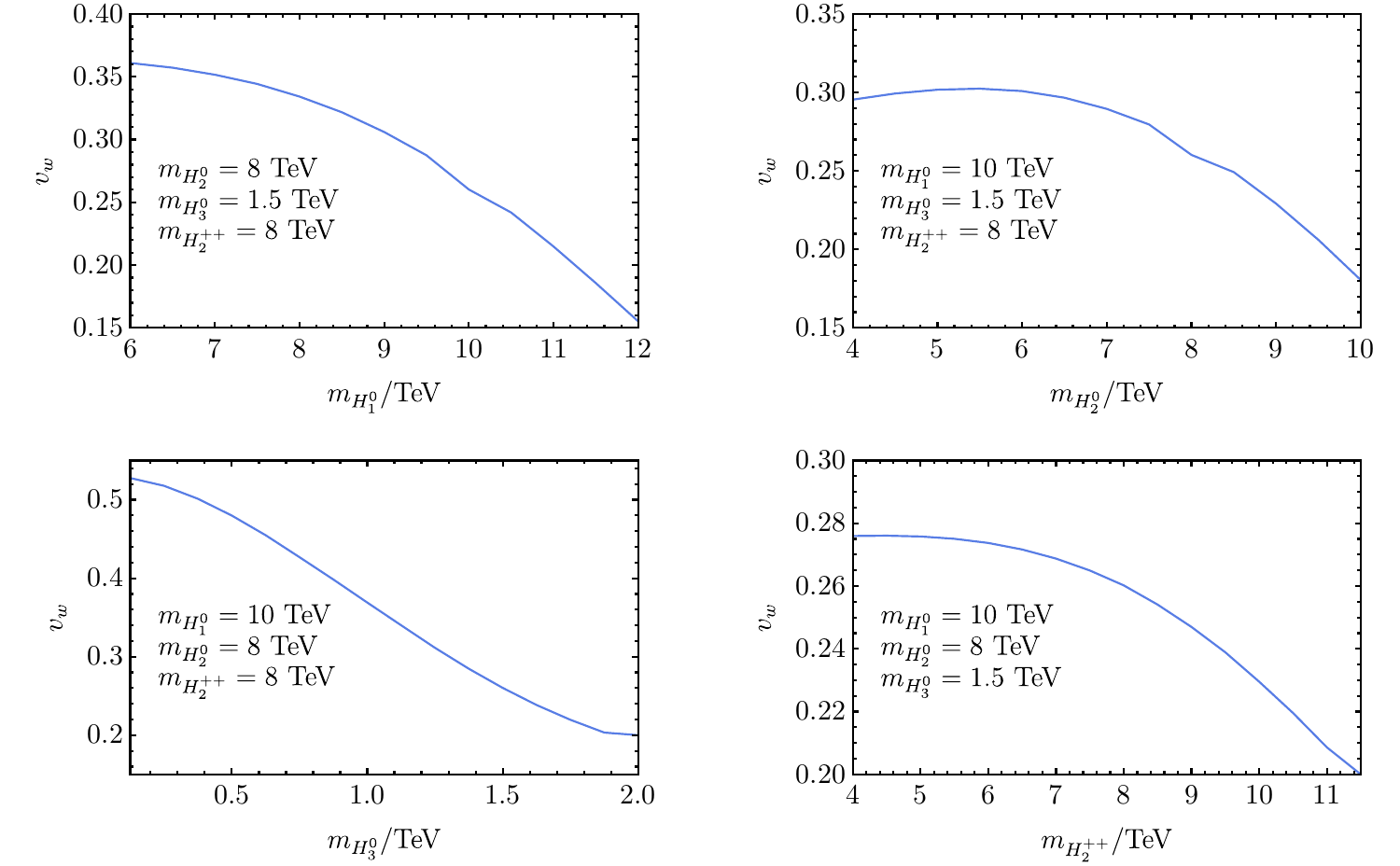}
		\caption{The bubble wall velocity $ v_w $ as a function of scalar particle masses are demonstrated.}
		\label{fig:vwmH}
	\end{figure}
	
	\begin{figure}[h!]
		\centering
		\includegraphics[width = 0.7\columnwidth]{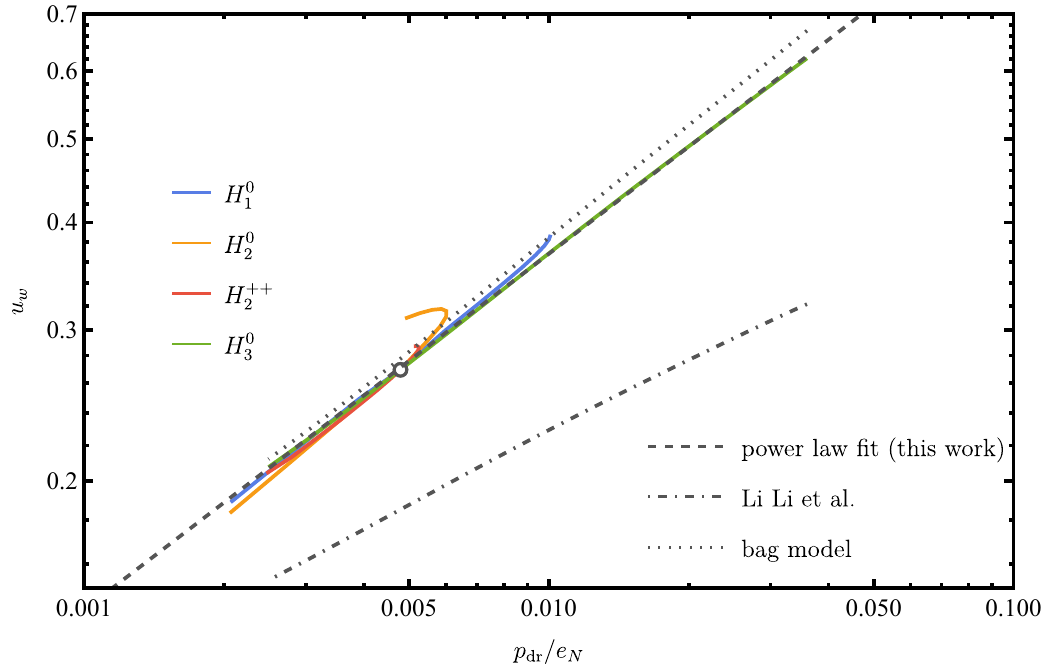}
		\caption{The $ u_w $ as a function of $ p_\mathrm{dr} / e_N $ is shown. Solid lines correspond to the cases where we vary masses of different particle species. The dashed line is a power law with slope 0.41. (Dash-)dotted line is the numerical (analytical) result given by Li Li et al. group's work \cite{cite:WSJresult}.}
		\label{fig:vwDp}
	\end{figure}
	
In \figurename{} \ref{fig:vwmH}, we demonstrate the bubble wall velocity $ v_w $ for different particle masses. We find that $ v_w $ does not always change monotonically with particle masses, because $ v_w $ is determined not only by a driven force due to the pressure difference, but also by the friction due to the interaction between particles and the scalar field.

It is useful to examine the pressure difference normalized by energy density, which is defined as follows
	\begin{equation}
	\label{eq:pdrdef}
		\frac{p_\mathrm{dr}}{e_N} \equiv \frac{p(\xi = 0) - p(\xi = 1)}{e(\xi = 1)},
	\end{equation}
	where $ p $ and $ e $ is read of from Eq. \eqref{eq:defthermalfunc}. In our work, when particle masses are large enough, both $ v_w $ and $ p_\mathrm{dr} / e_N $ tend to decrease. In such case, we found $ u_w $ and $ p_\mathrm{dr} / e_N $ follows power law, with the slope $ \simeq 0.41 $, where $ u_w \equiv v_w / \sqrt{1 - v_w^2} $, as shown in \figurename{} \ref{fig:vwDp}.
    
	As a comparison, we calculated $ u_w $ as a function of $ p_\mathrm{dr} / e_N $ and $ \alpha_N $ based on the bag-model. We use the method as Li Li et al. used in \cite{cite:WSJresult}. In that work, the authors treat $ p_\mathrm{dr} / w_N $ (or $ p_\mathrm{dr} / e_N $, equivalently) and $ \alpha_N $ as two known variables, and solve for $ \alpha_+ $ and $ \xi_w $ as unknown. In order to check the deviation between different method, we first solve for $ \xi_w $ using Eq. \eqref{eq:tzztrreq}, calculating $ p_\mathrm{dr} / w_N $ and $ \alpha_N $, then feed them into \cite{cite:WSJresult}'s formula
	\begin{equation}
	\label{eq:WSJformula}
	    \frac{p_\mathrm{dr}}{w_N} = 3 \alpha_N \qty(1 + \frac{9}{4} \alpha_N) \xi_{w, \mathrm{ana}}^2,
	\end{equation}
	and solve for $ \xi_{w, \mathrm{ana}} $, the result is shown by dash-dotted line in \figurename{} \ref{fig:vwDp}. Eq. \eqref{eq:WSJformula} is an analytical approximation based on the bag-model
	\begin{equation}
	    p_\pm = \frac{1}{3} a_\pm T^4 - V_\pm,\quad
	    e_\pm = a_\pm T^4 + V_\pm.
	\end{equation}
	Following the same procedure as in \cite{cite:WSJresult}, we also solve the exact $ \xi_{w, \mathrm{bag}} $ for the bag-model, using the same set of $ p_\mathrm{dr} / w_N $ and $ \alpha_N $ as input. The exact result for the bag-model is shown in \figurename{} \ref{fig:vwDp} with dotted line. For the case of a phase transition in radiation dominance epoch, with $ \alpha_N \lesssim 0.1 $, our result (dashed line) does not deviate much from the bag-model (dotted line). The reason is there are enough degree of freedom for passive species, the effective relativistic degree of freedom $ a_\pm $ and sound velocity $ c_{s,\pm} $ does not change much after bubble wall passes. In such case, the bag-model is still a good approximation for hydrodynamics when solving $ v_w $. However, in \cite{cite:WSJresult}, one knows $ p_\mathrm{dr} / e_N $ and $ \alpha_N $ \textit{a priori} to solve for $ \xi_w $, which is not true in our case. Thus solving $ \xi_w $ barely with knowledge of hydrodynamics is still impossible in our work.
	
	\begin{figure}[h!]
		\centering
		\includegraphics[width = 1\columnwidth]{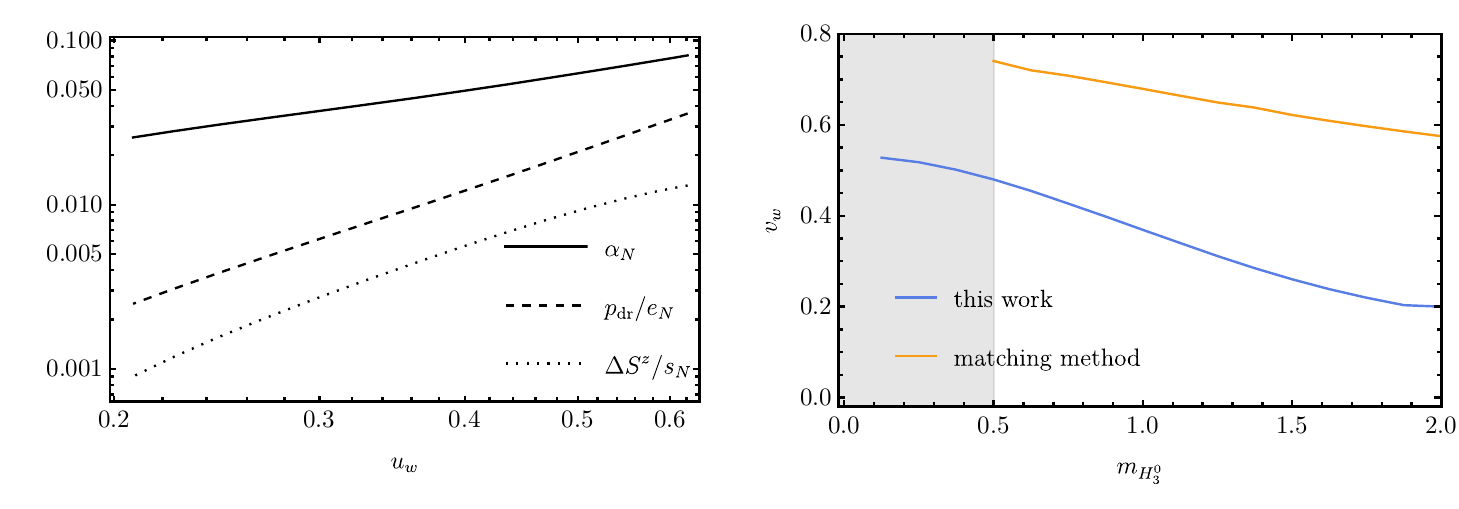}
		\caption{The entropy production rate $\Delta S^z/s_N$, $\alpha_N$, and $p_{dr}/e_N$, are shown as a function of $u_w$, with $ m_{H_3^0} $ varying from 0.125 TeV to 2 TeV, at the left panel. At the right panel, we compare the value of $v_w$ in our method with that of W.Y. Ai's matching method \cite{cite:AWYresult}. In the shaded region, Ai's method leads to runaway solutions.}
		\label{fig:AWY}
	\end{figure}
	
	In \cite{cite:AWYresult}, W.Y. Ai et al. provided a method to calculate $ \xi_w $ by adding an additional matching condition at bubble wall, i.e. one can assume
	\begin{equation}
	\label{eq:AWYformula}
	    S^z_\mathrm{LTE} = s_+ \gamma_+ v_+ = s_- \gamma_- v_-
	\end{equation}
	if the entropy difference is small enough, i.e. when $ \Delta s / s_+ \ll 1 $. We check the entropy production rate in our work, finding that $ \Delta S^z_\mathrm{LTE} / s_N $ follows almost the same power law as $ p_\mathrm{dr} / e_N $, up to some coefficient, as \figurename{} \ref{fig:AWY} shows in left panel. We also calculate $ \xi_w $ with combing Eq. \eqref{eq:AWYformula} and \eqref{eq:tzzt0zhydro}, and find the result is much larger, while the bubble wall will even runaway in some region, as shown in the right panel of \figurename{} \ref{fig:AWY}. Thus, entropy production rate can not be neglected, even if for small value.

	\section{Gravitational wave}
	\label{sec:GW}
	The stochastic gravitational waves generated by FOPT can be described in three parts \cite{cite:spectrum}: collision of bubble walls, sound waves(SW) and MHD turbulence. In this work, the bubble wall has reach its terminal velocity, thus the contribution from bubble collision can be neglected 
	\begin{equation}
		h^2 \Omega_\mathrm{GW} \simeq h^2 \Omega_\mathrm{SW}.
	\end{equation}
	The contribution from sound wave can be written as\cite{cite:wangzhiwei20}
	\begin{equation}
		h^2 \Omega_\mathrm{SW}(f) \simeq 2.65 \times 10^{-6} \qty(\frac{H_*}{\beta}) \qty(\frac{\kappa_v \alpha}{1 + \alpha})^2 \qty(\frac{100}{g_*})^{1 / 3} v_w \qty(\frac{f}{f_\mathrm{SW}})^3 \qty[\frac{7}{4 + 3 (f / f_\mathrm{SW})^2}]^{7 / 2} \Upsilon,
	\end{equation}
	where $ f $ is the frequency and the pivot frequency is given by
	\begin{equation}
		f_\mathrm{SW} \simeq 1.9 \times 10^{-2} \frac{1}{v_w} \qty(\frac{\beta}{H_*}) \qty(\frac{T_*}{100 \mathrm{GeV}}) \qty(\frac{g_*}{100})^{1 / 6} \mathrm{mHz}.
	\end{equation}
	The efficiency factor $ \kappa_v $ is the fraction of vacuum energy that is converted to bulk motion is calculated from \cite{cite:ErgBdg, cite:Hindmarsh15}
	\begin{equation}
		\label{eq:kv}
		\alpha \kappa_v = \frac{3}{e_N \xi_w^3} \int w(\xi)\, \gamma^2 v^2 \xi^2 \dd{\xi}, \qq{where}
		e_N = e_+(T_N).
	\end{equation}
	The suppression factor $ \Upsilon $ accounts the finite lifetime of sound wave in the medium, as suggested in \cite{cite:GuoHuaiKe21}
	\begin{equation}
		\Upsilon = 1 - 1 / \sqrt{1 + 2 \tau_\mathrm{SW} H_*},
	\end{equation}
	where lifetime $ \tau_\mathrm{SW} $ can be estimated as \cite{cite:Ellis20}
	\begin{equation}
		\tau_\mathrm{SW} H_* \sim H_* R_* / \bar{U}_f,
	\end{equation}
	with $ \bar{U}_f $ the root-mean-square of the bulk fluid velocity \cite{cite:Hindmarsh15}
	\begin{equation}
		\bar{U}_f^2 = \frac{3}{w_N (1 + \alpha) \xi_w^3} \int w(\xi)\, \gamma^2 v^2 \xi^2 \dd{\xi} = \frac{e_N}{w_N} \frac{\kappa_v \alpha}{1 + \alpha}, \qq{where}
		w_N = w_+(T_N).
	\end{equation}
	
	\begin{figure}[h!]
		\centering
		\includegraphics[width = 0.7\columnwidth]{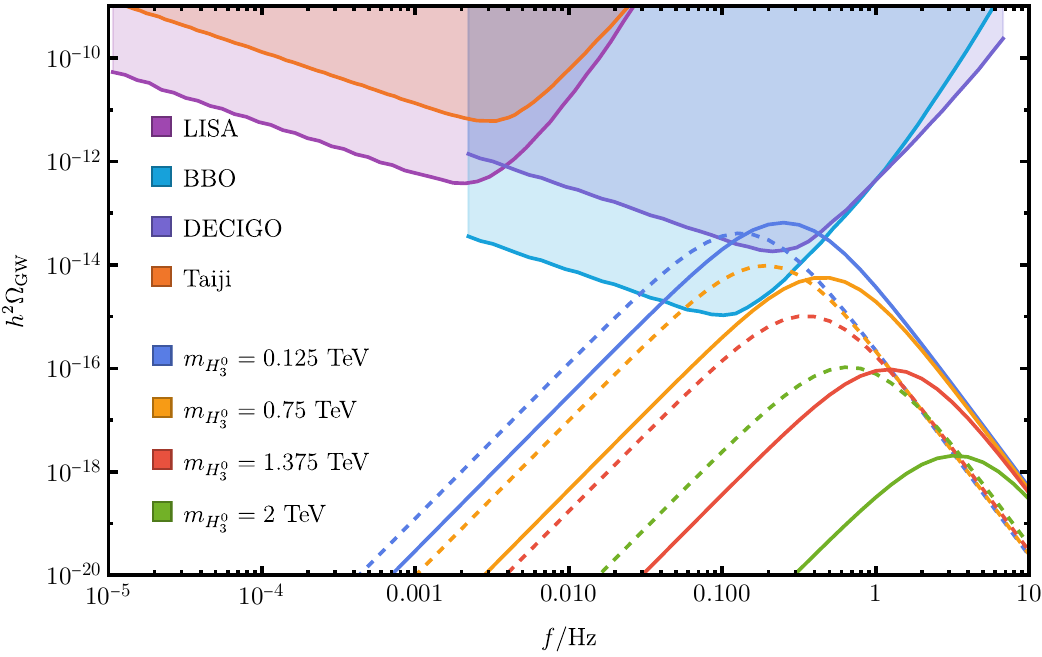}
		\caption{Predicted GW spectra for $ v_w $ calculated from Eq. \eqref{eq:tzztrreq} are shown as solid curves, while the dashed curves denote the results by taking a constant $ v_w = 1 $. Four different values of $ m_{H_3^0} $, i.e. 0.125 TeV, 0.75 TeV, 0.375 TeV, 2 TeV, are taken for each pair of curves. For each pair of solid and dashed curves of the same color, model parameters are the same. Sensitivity curves for detectors, like LISA, BBO, DECIGO and Taiji, are shown in shaded regions.}
		\label{fig:GWmH3}
	\end{figure}
	\begin{figure}[h!]
		\centering
		\includegraphics[width = 0.7\columnwidth]{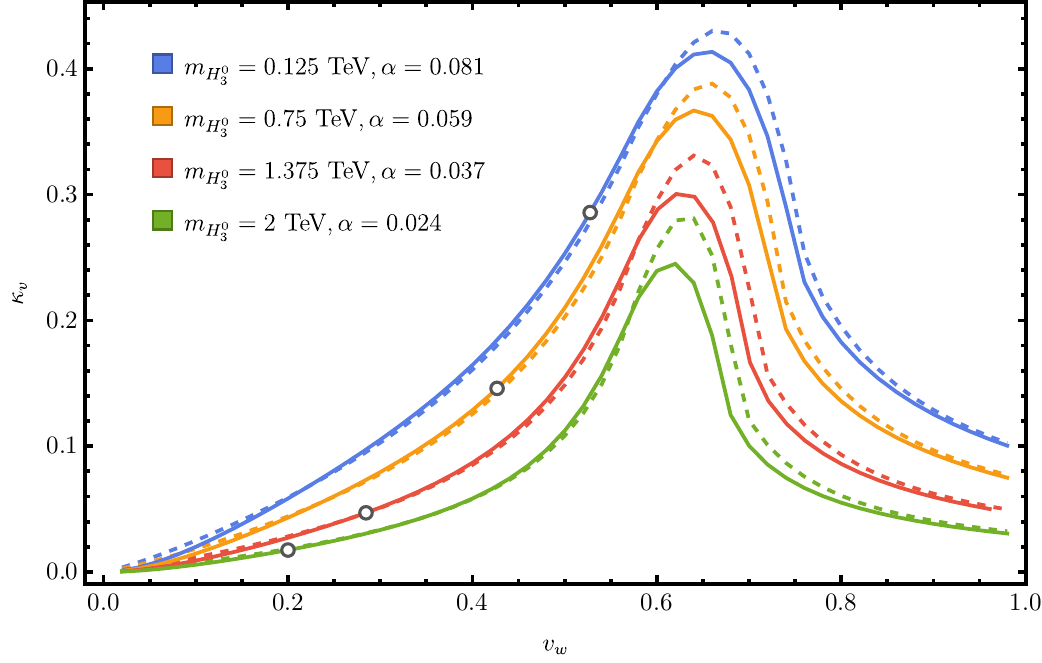}
		\caption{We show the value of $ \kappa_v $ as a function of $ v_w $ for different $ m_{H_3^0} $. Solid lines are calculated from Eq. \eqref{eq:kv}, with $ v_w $ as a free parameter, dashed lines are calculated from numerical fitted formula given in \cite{cite:ErgBdg}. Open circles correspond to $ v_w $ calculated from Eq. \eqref{eq:tzztrreq}.}
		\label{fig:kvvw}
	\end{figure}
	We calculated the GW spectra, the result are shown in \figurename{} \ref{fig:GWmH3}. The solid curves are the predicted spectra for $ H_3^0 $ mass from 0.125 TeV to 2 TeV. In the earlier work \cite{cite:mingqiu}, a constant $ v_w = 1 $ was taken. The predicted spectra of the earlier work are given as dashed curves. It is observed that the smaller the parameter $m_{H_3^0}$, the larger is the amplitude of GW. It is found that when $v_w$ is determined dynamically, the GW spectra can have higher peaks when compared with the case if we take $ v_w = 1 $. The enhancement can be attributed to the fact that the $ \kappa_v $ factor takes its maximum value when $ c_s \lesssim v_w $. The relation of $ \kappa_v $ versus $ v_w $ for different $ m_{H_3^0} $are shown in \figurename{} \ref{fig:kvvw}. Compared with the case $ v_w = 1 $, the value of $ \kappa_v $ can vary in drastically when the dynamic $ v_w $ from Eq. \eqref{eq:tzztrreq} are adopted. For the best case in this work, $ \kappa_v^2 $ can be enhanced by $ 1 \sim 2 $ order of magnitude due to this effect. However, since $ \Upsilon $ factor dependent on $ v_w $ and $ \kappa_v $, this enhancement becomes less efficient in the determination of the peak amplitudes of GW spectra.
	
	\section{Discussion and Conclusion}
	\label{sec:dnc}
	In this work we studied the bubble wall velocity determined by the phase transition dynamics in the MLRSM. We calculated $ \alpha $, $ \beta / H_* $ and $ v_w $ in the first order phase transition parameter space. \figurename{} \ref{fig:alphabeta} demonstrates that the parameter $ m_{H_3^0} $ is the most important parameter in determining the phase transition parameters $ \alpha $ and $ \beta / H_* $.
	As shown in \figurename{} \ref{fig:vwmH}, it is found that the bubble wall velocity can be determined in the range  $  0.2 < v_w <  0.5 $ in the MLRSM, and its value tends to decrease when the masses of scalars increase. In the model parameter regions with a very large scalar mass, the relation between bubble wall velocity and the pressure difference can be described by a power law in the form $ \gamma_w v_w \propto [(p_C - p_A) / e_A]^{0.41} $, as shown in \figurename{} \ref{fig:vwDp}. Compared with previous work \cite{cite:mingqiu}, where $ v_w = 1 $ was taken, we found that the peak amplitudes of GW spectra can be enhanced and the peak frequencies can be shifted to high.
	
\begin{figure}[h!]
		\centering
		\includegraphics[width = 0.7\columnwidth]{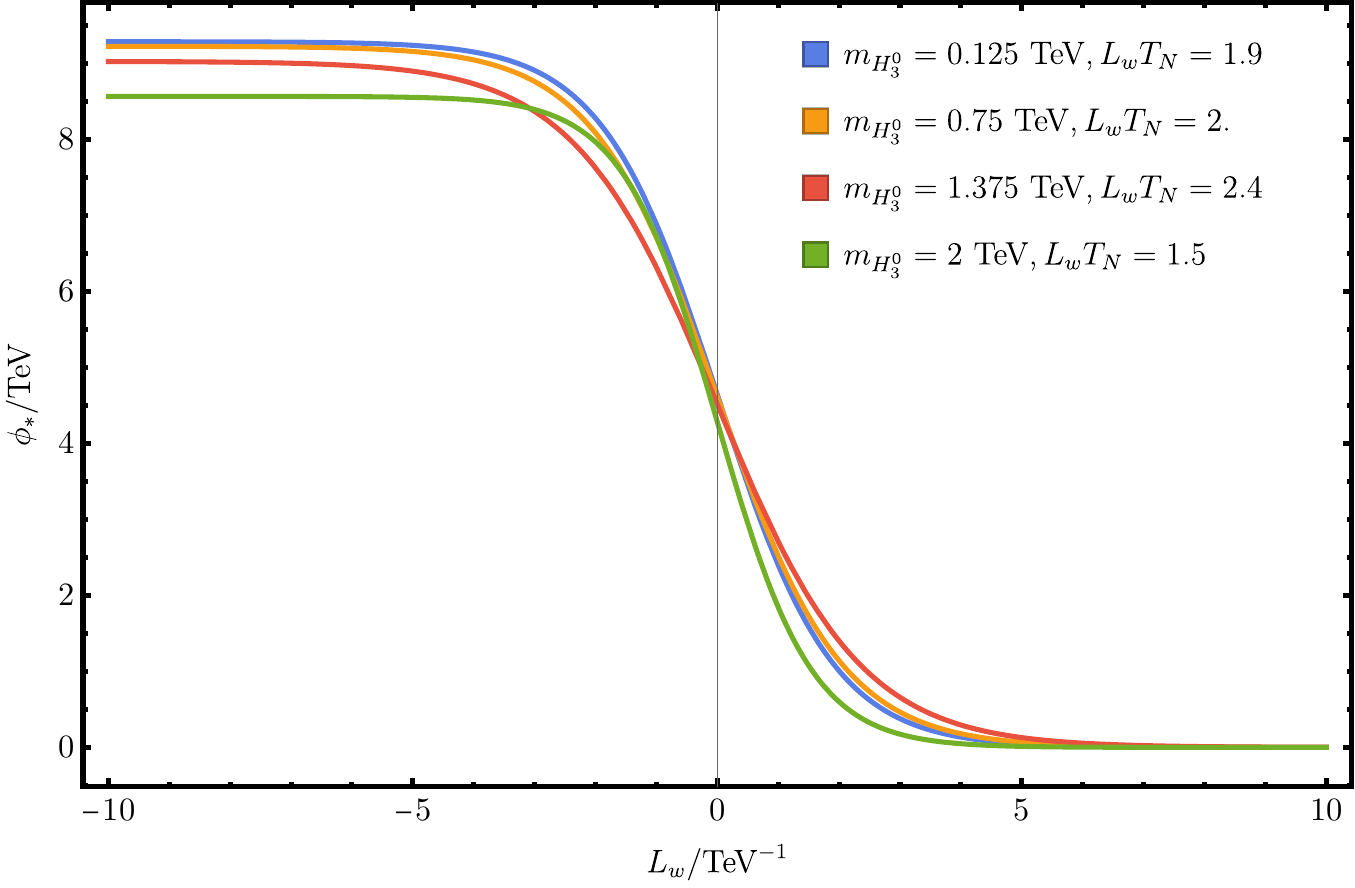}
		\caption{The kink solutions of scalar field $\phi$ are demonstrated, and the bubble wall thicknesses $ L_w \sim 2 \, T_N^{-1} $ are shown.}
		\label{fig:tanh}
	\end{figure}

    In order to check the validity of Eq \eqref{eq:Lwsmall}, we also calculated $ L_w $. We adopt a $ \tanh $ ansatz for profile of $ \phi(z) $
	\begin{equation}
		\label{eq:scalarAnsatz}
		\phi(L_w; z) = \frac{\phi_*}{2} \qty(1 + \tanh(\frac{z}{L_w})),
	\end{equation}
	where $ \phi_* $ is the local minimum of $ T^{zz}(T, \phi, v_w) $. We can define an action $ F(v) $ according to Eq. \eqref{eq:scalarEoM}
	\begin{equation}
		\label{eq:zAction}
		F = \int \dd{z} \frac{1}{2} (\pd_z \phi)^2 + \qty[V_\mathrm{vac}(\phi) + \sum_i \int \frac{p_z^2 \dd[3]p}{(2 \pi)^3 E_i} f_i(\vb p, \vb x))]_{\phi = 0}^{\phi = \phi(z)}.
	\end{equation}
	Inserting Eq. \eqref{eq:scalarAnsatz} into Eq. \eqref{eq:zAction}, then we can solve for $ L_w $ by minimizing $ F $. The results are shown in \figurename{} \ref{fig:tanh}. We found that the wall thickness as $ \order{2} $. Since we have $ v_w = 0.2 \sim 0.5 $ as our result, the corresponding $ \gamma_w^{-1} = 0.8 \sim 1 $. Thus Eq. \eqref{eq:Lwsmall} is valid in our work.
	
	In order to determine the bubble wall velocity, we have combined the energy momentum conservation relations in the free particle approximation and LTE approximation, as given in Eq. (\ref{eq:tzzt0zeq}).
	Naively speaking, in the LTE approximation, one might expect that $ p_- > p_+ $, since the bubble is pushed to expand. However, it turns out that this is not necessarily to be true. For example, one can easily check this by plugging the first line in Eq. \eqref{eq:tzzt0zhydro} to the second line, one obtain
	\begin{equation}
    T^{0r}_\mathrm{LTE} = \frac{p_+ - p_-}{v_- - v_+},\qq{and}
    T^{rr}_\mathrm{LTE} = \frac{p_+ v_- - p_- v_+}{v_- - v_+}.
	\end{equation}
	Since $ T^{0r}_\mathrm{LTE} $ should always be positive, there should be a sign flip for $ p_+ - p_- $ between the detonation mode and the deflagration mode. This behavior can be also seen from the lower left panel in \figurename{} \ref{fig:xiv}, as $ p \propto T^4 $ for $ \alpha_N \ll 1 $. However, the quantity $ p_\mathrm{dr}/e_N $ we have used in Eq. \eqref{eq:pdrdef} is well defined, which is positive both in the detonation mode and the deflagration mode.

\begin{figure}[h!]
		\centering
		\includegraphics[width = 0.7\columnwidth]{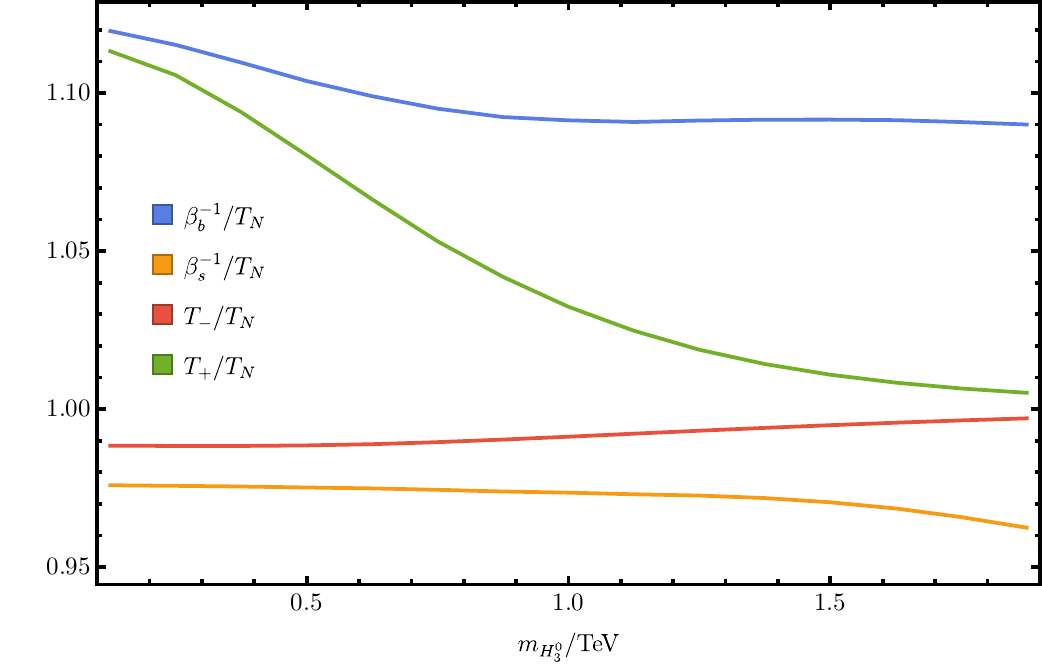}
		\caption{The solutions of $\beta_{b,s}^{-1} /T_N$ and $T_\pm/T_N$ are demonstrated.}
		\label{fig:tttt}
	\end{figure}

	In free particle approximation, we have introduced two parameters $ \beta_{b,s}^{-1} $ for all particles, based on the ansatz that the incident particles are in thermal equilibrium. Together with $ v_w $, there are three free parameters to solve the matching equations given in Eq. (\ref{eq:tzzt0zeq}). It is natural to expect that $ \beta_{b,s}^{-1} $ should have similar behavior with $ T_\mp $, where $ T_\mp $ denote the temperatures on the two sides of bubble wall in the LTE approximation. It turns out that our numerical solutions do not support this expectation. In \figurename{} \ref{fig:tttt}, we show the solutions of $\beta_{b,s}^{-1} $, together with $ T_\pm $, normalized by $ T_N $. Our results demonstrate that there are two relations: 1)  $ \beta_b^{-1} > T_N > \beta_s^{-1} $ and 2) $ T_+ > T_N > T_- $, which do not agree with this expectation. 
	
	The underlying reason might be traced to our ansatz which has neglected the collision terms in the Boltzman equations. The inconsistency might be resolved after taking into account the collisions between the incident particles with transmitted/reflected particles, within the distance $ \lambda_\mathrm{mfp} $ away from the bubble wall. Generally speaking, the collision processes inside and outside the bubble wall can be different when the FOPT occurs. For example, massive particles decaying can only happen in the broken phase. To treat the effects of collision terms in an appropriate way, we should assume that different particles should have different $\beta_b$ and $\beta_s$ and should solve the coupled  Boltzmann equation numerically \cite{cite:MoPr95, cite:newnewform}, which can be done in our future work. Furthermore, the release of vacuum energy during the FOPT might reheat the particles on and inside the wall. Other physics processes, like the decay of active particles into the passive SM particles, can increase the temperature of plasma and consequently can affect the distribution functions of all particles inside the bubble wall. All these effects should be handled appropriately. Obviously, these effects can change the relation between $ v_w $ and $ \Delta T^{zz}_{i_\pm} $ and deserve our future study. 

\appendix
\section{Particle masses and thermal corrections}
\label{sec:mass}

In this appendix, we provide the mass formula for scalar and vector bosons and fermions which can contribute to the effective potential. 
	
	For real components of neutral scalar particles, in the basis of $ \Re \{ \phi_1^0, \phi_2^0, \Delta_R^0, \Delta_L^0 \} $, the mass matrix elements are
	\begin{equation}
	    \begin{split}
	        M_{11}^{\Re}(T, v_R) &= - \mu_1^2 + \alpha_1 / 2 \, v_R^2 + \Pi_{11}(T),\\
	        M_{22}^{\Re}(T, v_R) &= - \mu_1^2 + (\alpha_1 + \alpha_3) / 2 \, v_R^2 + \Pi_{11}(T),\\
	        \Pi_{11}(T) &= T^2 / 24 \, (9 / 2 g_L^2 + 9 / 2 g_R^2 + 20 \lambda_1 + 8 \lambda_3 + 12 \alpha_1 + 6 \alpha_3 + 6 y_t^2 + 6 y_b^2),\\
	        M_{12}^{\Re}(T, v_R) &= - 2 \mu_2^2 + \alpha_3 / 2 \, v_R^2 + \Pi_{12}(T),\\
	        \Pi_{12}(T) &= T (\alpha_2 + \lambda_4 + y_t y_b),\\
	        M_{33}^{\Re}(T, v_R) &= - \mu_3^2 + 3 \rho_1 v_R^2 + \Pi_{33}(T),\\
	        \Pi_{33}(T) &= T^2 / 24 \, (
	        12 g_L^2 + 6 g_{B-L}^2 + 16 \rho_1 + 8 \rho_2 + 6 \rho_3 + 8 \alpha_1 + 4 \alpha_3 + 12 y_N^2),\\
	        M_{44}^{\Re}(T, v_R) &= - \mu_3^2 + \rho_3 / 2 \, v_R^2 + \Pi_{44}(T),\\
	        \Pi_{44}(T) &= T^2 / 24 \, (
	        12 g_R^2 + 6 g_{B-L}^2 + 16 \rho_1 + 8 \rho_2 + 6 \rho_3 + 8 \alpha_1 + 4 \alpha_3 + 12 y_N^2).
	    \end{split}
	\end{equation}
	For imaginary components of neutral scalar particles, in the basis of $ \Im \{ \phi_1^0, \phi_2^0, \Delta_R^0, \Delta_L^0 \} $, the mass matrix elements are
	\begin{equation}
	    \begin{split}
	        M_{11}^{\Im}(T, v_R) &= M_{11}^{\Re}(T, v_R),\quad
	        M_{22}^{\Im}(T, v_R) = M_{22}^{\Re}(T, v_R),\quad
	        M_{12}^{\Im}(T, v_R) = - M_{12}^{\Re}(T, v_R),\\
	        M_{33}^{\Im}(T, v_R) &= - \mu_3^2 + \rho_1 v_R^2 + \Pi_{33}(T),\quad
	        M_{44}^{\Im}(T, v_R) = M_{44}^{\Re}(T, v_R).
	    \end{split}
	\end{equation}
	For charged scalar particles, in the basis of $ \{ \phi_1^\pm, \phi_2^\pm, \Delta_R^\pm, \Delta_L^\pm \} $, the mass matrix elements are
	\begin{equation}
	    \begin{split}
	        M_{11}^{\pm}(T, v_R) &= M_{11}^{\Re}(T, v_R),\quad
	        M_{22}^{\pm}(T, v_R) = M_{22}^{\Re}(T, v_R),\quad
	        M_{12}^{\pm}(T, v_R) = + M_{12}^{\Re}(T, v_R),\\
	        M_{33}^{\pm}(T, v_R) &=  - \mu_3^2 + \rho_1 \, v_R^2 + \Pi_{33}(T) 
	        = M_{33}^{\Im}(T, v_R),\quad
	        M_{44}^{\pm}(T, v_R) = M_{44}^{\Re}(T, v_R),
	    \end{split}
	\end{equation}
	For double charged scalar particles, in the basis of $ \{ \Delta_R^{\pm\pm}, \Delta_L^{\pm\pm} \} $, the mass matrix elements are
	\begin{equation}
	    \begin{split}
	        M_{11}^{\pm\pm}(T, v_R) &= - \mu_3^2 + (\rho_1 + 2 \rho_2) \, v_R^2 + \Pi_{33}(T),\quad
	        M_{22}^{\pm\pm}(T, v_R) = M_{44}^{\pm}(T, v_R).
	    \end{split}
	\end{equation}
	By diagonalizing $ M^{\Re}, M^{\Im} $ and $ M^{\pm} $, one find there is one eigen value for each matrix that barely depend on $ v_R $. These state corresponding to the Higgs doublet in SM. Since we take $ \alpha_1 = \alpha_2 = 0 $, this doublet barely contribute to the $ v_R $ phase transition, thus we don't include them in Eq. \eqref{eq:Veff} and following calculation. There are also degeneracies in remaining eigen values in mass matrices, only 5 masses are unique, to be explicit, they are
	\begin{equation}
	\begin{split}
        m^2_{H_1^0} &= m^2_{A_1^0} = m^2_{H_1^\pm} \in M_{\Phi},\quad
        m^2_{H_2^0} = m^2_{A_2^0} = m^2_{H_2^\pm} = m^2_{H_1^{\pm\pm}} \in M_{\Delta_L},\\
        m^2_{\eta^0} &= m^2_{\eta^\pm} \in M_{\Delta_R},\quad
        m^2_{H_3^0} \in M_{\Delta_R},\quad
        m^2_{H_2^{\pm\pm}} \in M_{\Delta_R}.
	\end{split}\label{massrel}
	\end{equation}
	Among these states, $ \eta^0 $ and $ \eta^\pm $ are corresponding to the pseudo-Nambu-Goldstone bosons in the $ SU(2)_R $ symmetry breaking, these states are replaced by longitudinal gauge bosons in the following calculation.
	
	For gauge bosons, $ \vec W_R $ and $ B_{B-L} $ becomes massive $ W^{\pm}_R $ and $ Z^0_R $, together with a massless $ B_Y $ after symmetry breaking, while leaving $ \vec W_L $ unchanged. The masses for gauge bosons are
	\begin{equation}
	    m^2_{W^\pm_R} = g_R^2 / 2 \, v_R^2,\quad
	    m^2_{Z^0_R} = \qty(g_R^2 + g_{B-L}^2) \, v_R^2.
	\end{equation}
	
	For fermions, we consider three generations of right handed neutrino to have the same mass
	\begin{equation}
	    m_{N_R} = \sqrt2 y_N \, v_R.
	\end{equation}
	\begin{acknowledgements}
We are grateful to discussions with Yidian Chen, Anping Huang, Mingqiu Li and Shaojiang Wang.
This work is supported in part by the National Natural Science Foundation of China (NSFC) Grant  Nos: 12235016, 12221005, and supported by the Strategic Priority Research Program of Chinese Academy of Sciences under Grant No XDB34030000, and the Fundamental Research Funds for the Central Universities.
\end{acknowledgements}
	
\bibliography{reference}
\end{document}